\RequirePackage[T1]{fontenc}
\documentclass[12pt]{article}

\usepackage{times}

\pdfoutput=1
\usepackage[height=8.85in,width=6.45in]{geometry}

\usepackage[utf8]{inputenc}
\usepackage{amsmath}
\usepackage{amssymb}
\usepackage{mathtools}
\numberwithin{equation}{section}
\usepackage{slashed}
\usepackage{braket}
\usepackage[svgnames,psnames]{xcolor}
\usepackage[colorlinks,citecolor=DarkGreen,linkcolor=FireBrick,linktocpage,unicode]{hyperref}
\usepackage{cite}
\usepackage{graphicx}
\usepackage{tikz}
\usepackage{tikz-cd}
\usepackage{spectralsequences}
\usepackage[whole]{bxcjkjatype} 
\usepackage[bottom]{footmisc}

\usepackage{courier}
\usepackage{bm}
\usepackage{subfig}
\usepackage{dashbox}

\usepackage{mdframed}

\renewenvironment{figure}[1][]{
  \begin{originalfigure}[#1]
    \begin{mdframed}[linecolor=black!0,backgroundcolor=black!1]
}{
    \end{mdframed}
  \end{originalfigure}
}
\renewenvironment{table}[1][]{
  \begin{originaltable}[#1]
    \begin{mdframed}[linecolor=black!0,backgroundcolor=black!1]
}{
    \end{mdframed}
  \end{originaltable}
}%% Comment

\usepackage{colortbl}
\definecolor{lightyellow}{rgb}{1.0, 0.95, 0.7}
\definecolor{lightblue}{rgb}{0.7, 0.9, 1.0}
\definecolor{lightpink}{rgb}{1.0, 0.85, 0.95}
\definecolor{lightgreen}{rgb}{0.7, 1.0, 0.4}
\definecolor{lightorange}{rgb}{1.0, 0.85, 0.6}
\definecolor{lightpurple}{rgb}{0.95, 0.9, 1.}
\definecolor{blue}{rgb}{0.0, 0.4, 1.0}
\definecolor{Blue}{rgb}{0,0,1}
\definecolor{darkgreen}{rgb}{0.,0.6,0.}

\newcommand*{\red}[1]{\textcolor{red}{#1}}
\newcommand*{\Blue}[1]{\textcolor{Blue}{#1}}
\newcommand*{\black}[1]{\textcolor{black}{#1}}

\newcommand*{\textarrow}[1]{\xrightarrow{\hspace{2mm}\text{#1}\hspace{2mm}}}

\newcommand*{\CP}{\mathbb{CP}}
\newcommand*{\HP}{\mathbb{HP}}
\newcommand*{\bZ}{\mathbb{Z}}
\newcommand*{\bR}{\mathbb{R}}
\newcommand*{\dint}{\displaystyle\int}
\newcommand*{\cI}{\mathcal{I}}
\newcommand*{\cA}{\mathcal{A}}

\newcommand*{\cN}{\mathcal{N}}
\newcommand*{\cP}{\mathcal{P}}

\def\Nequals#1{$\mathcal{N}{=}#1$}
\def\bQ{\mathbb{Q}}

\def\tr{\mathop{\mathrm{Tr}}\nolimits}
\def\Ext{\mathop{\mathrm{Ext}}\nolimits}
\def\Hom{\mathop{\mathrm{Hom}}\nolimits}
\def\Tors{\mathop{\mathrm{Tors}}\nolimits}
\def\Free{\mathop{\mathrm{Free}}\nolimits}
\def\Inv{\mathop{\mathrm{Inv}}\nolimits}

\def\ch{\mathop{\mathrm{ch}}\nolimits}
\usepackage{arydshln}

\def\boo{0.0}

\def\xlattice#1#2#3{
\begin{tikzpicture}[scale=.5]
\filldraw[color=black!5!white](-.5,-.5) rectangle (1.5,1.5);
\draw[->] (-1,0) -- (2,0);
\draw[->] (0,-1) -- (0,2);
\foreach \x in {0,1} {
	\foreach \y in {0,1}{
		\pgfmathsetmacro\a{mod(#1 * \x - #2 * \y,2)}
		\ifx\a\boo
			\filldraw[color=#3] (\x,\y) circle (.5em);
		\else
			\filldraw[fill=white,draw=gray] (\x,\y) circle (.5em);
		\fi
	}
}
\end{tikzpicture}
}

\begin{document}

\begin{titlepage}

\begin{flushright}
%IPMU-**-****
\end{flushright}

\vskip 3cm

\begin{center}

{\Large \bfseries 
	Some comments on $6d$ global gauge anomalies
}

\vskip 1cm
Yasunori Lee 
and Yuji Tachikawa
\vskip 1cm

\begin{tabular}{ll}
 & Kavli Institute for the Physics and Mathematics of the Universe (WPI), \\
& University of Tokyo,  Kashiwa, Chiba 277-8583, Japan
\end{tabular}

\vskip 1cm

\end{center}

\noindent 
Global gauge anomalies in $6d$
associated with non-trivial homotopy groups $\pi_6(G)$ for $G=SU(2)$, $SU(3)$, and $G_2$
were computed and utilized in the past.
In the modern bordism point of view of anomalies, however, 
they come from the bordism groups $\Omega^\text{spin}_7(BG)$, 
which are in fact trivial and therefore preclude their existence.
Instead, it was noticed that a proper treatment of the $6d$ Green-Schwarz mechanism reproduces the same anomaly cancellation conditions derived from $\pi_6(G)$.
In this paper, we revisit and clarify the relation between these two different approaches.

\vskip 1cm

\begin{center}
\itshape
--- dedicated to the memory of the late Professor Tohru Eguchi ---
\end{center}

\end{titlepage}

\setcounter{tocdepth}{2}
\tableofcontents

%\bigskip
\newpage

\section{Introduction and summary}

Given a gauge theory, its gauge anomaly must be canceled as a whole in order for the theory to be consistent.
This imposes non-trivial constraints on the possible matter content.
One notable subtlety here is that,
even if the \emph{perturbative} anomaly cancellation is achieved, the theory may still suffer from \emph{global} anomaly,
coming from a global gauge transformation corresponding to non-trivial elements of $\pi_d(G)$,
where $d$ is the spacetime dimension and $G$ is the gauge group.
It was first pointed out in \cite{Witten:1982fp} that such a situation indeed arises for $4d$ $SU(2)$ gauge theory with a Weyl fermion in the doublet.

In $6d$ the situation is more subtle, since the anomaly cancellation often involves chiral 2-form fields through the Green-Schwarz mechanism.
The cancellation condition can nonetheless be derived,
by embedding $G$ in a larger group $\tilde G$ whose global anomaly is absent.
This approach was originally developed by Elitzur and Nair \cite{Elitzur:1984kr}, and
the extension to $6d$ with the Green-Schwarz mechanism was done 
in \cite{Tosa:1989qm,Bershadsky:1997sb,Suzuki:2005vu}.\footnote{%
There are many other works in the late 80s where the analysis in $6d$ was done without the Green-Schwarz mechanism.
We do not cite them here; interested readers can find them by looking up papers citing \cite{Elitzur:1984kr} in the INSPIRE-HEP database.
}
The list of simply-connected simple Lie groups with non-trivial $\pi_6(G)$ is\footnote{%
$\pi_6(G)$ for classical groups were computed in \cite[Sec.~19]{BorelSerre}.
$\pi_6(G)$ for $G_2$ and $F_4$ were considered as known in the review article \cite{borel1955} and were attributed to H.~Toda; a derivation can be found in \cite{mimura1967}, in which $\pi_i(G_2)$ and $\pi_i(F_4)$ were completely determined up to $i=21$.
$\pi_6(G)$ for $E_{6,7,8}$ were determined to be trivial in \cite[Theorem V]{BottSamelson}.
$\pi_6(G)$ for $G=SU(2)=S^3$ belongs more properly to the unstable homotopy groups of spheres and was computed by many people, see the footnote 7 of \cite[Sec.~19]{BorelSerre} and the comments in \cite[p.~428]{borel1955}.
} 
\begin{equation}
	\begin{array}{lcl}
		\pi_6(SU(2)) & = & \bZ_{12},\\
		\pi_6(SU(3)) & = & \bZ_6,\\
		\pi_6(G_2) & = & \bZ_3,\\
	\end{array}
\end{equation}
and the references above found a mod-12, mod-6, mod-3 condition for $G=SU(2)$, $SU(3)$, and $G_2$, respectively.
In \cite{Bershadsky:1997sb}, it was indeed found that the F-theory compactification to $6d$ only produces gauge theories which are free from the global gauge anomalies. 

From the modern point of view, 
the anomaly of a theory $Q$ in $d$ spacetime dimensions is described via 
its anomaly theory $A(Q)$ in $(d+1)$ dimensions,
which hosts the original theory $Q$ on its boundary.
The anomalous phase associated to a gauge transformation $g:S^d\to G$ 
is now re-interpreted as the partition function \begin{equation}
Z_{A(Q)} [S^{d+1},P_g]
\end{equation}
of the anomaly theory 
on $S^{d+1}$ equipped with the $G$ gauge field $P_g$ obtained by performing gauge transformation $g$ at the equator $S^d\subset S^{d+1}$.

More generally,  the possible anomalies of a symmetry group $G$ 
are characterized by $\Inv_{\text{spin}}^{d+1}(BG)$,
the group formed by deformation classes of $(d+1)$-dimensional invertible phases with $G$ symmetry.
This group fits in the following short exact sequence
\begin{equation}
	0
	\longrightarrow
	\Ext_\bZ(\Omega^{\text{spin}}_{d+1}(BG),\bZ)
	\longrightarrow
	\Inv_{\text{spin}}^{d+1}(BG)
	\longrightarrow
	\Hom_\bZ(\Omega^{\text{spin}}_{d+2}(BG),\bZ)
	\longrightarrow
	0
	\label{AndersonDual}
\end{equation}
where 
$\Omega^\text{spin}_d(BG)$ is the bordism group of $d$-dimensional spin manifolds equipped with $G$ bundle \cite{Freed:2016rqq}.
In particular, the information on global anomalies is encoded in the part
\begin{equation}
\Ext_\bZ(\Omega^{\text{spin}}_{d+1}(BG),\bZ) = 
\Hom(\Tors  \Omega^{\text{spin}}_{d+1} (BG),U(1)),
\end{equation}
while the information on the anomaly polynomial is encoded in the part \begin{equation}
\Hom_\bZ(\Omega^{\text{spin}}_{d+2}(BG),\bZ)
= \Hom_\bZ(\Free \Omega^{\text{spin}}_{d+2}(BG),\bZ).
\end{equation} 
For $4d$ theory with $G=SU(2)$ symmetry, the group $\Omega^\text{spin}_5(BSU(2))$ is indeed $\bZ_2$,
and is generated by $(S^5,P_{[g]})$ where $g:S^4\to SU(2)$ belongs to the generator of $\pi_4(SU(2))=\bZ_2$.
This corresponds to Witten's original global anomaly.

When we apply this argument to $6d$, we encounter an immediate puzzle.
Namely, for $G=SU(2)$, $SU(3)$, and $G_2$ for which $\pi_6(G)$ is non-trivial, we have
\begin{equation}
	\begin{array}{lcc}
		\Omega^{\text{spin}}_{7}(BSU(2)) & = & 0,\\
		\Omega^{\text{spin}}_{7}(BSU(3)) & = & 0,\\
		\Omega^{\text{spin}}_{7}(BG_2) & = & 0,\\
	\end{array}
\end{equation}
and in particular the configuration $(S^7, P_{[g]})$ for $[g]\in \pi_6(G)$ is null-bordant.
Therefore, the partition function of the anomaly theory on this background should be automatically trivial when the anomaly polynomial is canceled,
and there should be no global anomalies at all for the gauge groups $G=SU(2)$, $SU(3)$, and $G_2$.
But the old computations indeed found non-trivial anomalous phases,
so one is led to wonder whether they were legitimate to start with.

The way out is suggested by another set of observations made more recently, in \cite[Sec.~3.1.1]{Ohmori:2014kda} and
in \cite{Monnier:2018nfs,Monnier:2018cfa}.
Namely, it was observed there 
that the mod-12, mod-6, mod-3 conditions for $G=SU(2)$, $SU(3)$, $G_2$ theories can be derived by demanding that 
the factorized anomaly polynomial of the fermions can actually be canceled by a properly constructed Green-Schwarz term. 
For example, the non-(purely-)gravitational part of the anomaly polynomial\footnote{%
In this paper, we denote by $\cI$ the anomaly polynomial of a theory, 
and by $\tilde\cI$ the part of the anomaly polynomial not purely formed by the Pontrjagin classes $p_i(R)$ of the spacetime.
} of $n$ fermions in $\mathbf{3}$ of $SU(3)$ is \begin{equation}
\tilde\cI_\text{fermion}= \frac{n}{6} \cdot \frac12\cdot c_2(F) \left(c_2(F) + \frac{p_1(R)}{2}\right)\label{nnn}
\end{equation} and it is always factorized, which is a necessary condition for the Green-Schwarz mechanism to be applicable.
But it is not a sufficient condition.
The instanton configurations of $G$ gauge field correspond to strings charged under 2-form fields,
and their self Dirac pairing is given by $n/6$, which needs to be an integer.
This shows that $n$ needs to be divisible by $6$ \cite[Sec.~3.1.1]{Ohmori:2014kda}.
Furthermore, when this is the case, one can actually construct a theory of 2-form fields with this anomaly,
thanks to the series of works by Monnier and his collaborators  \cite{Monnier:2010ww,Monnier:2011mv,Monnier:2011rk,Monnier:2012xd,Monnier:2013kna,Monnier:2013rpa,Monnier:2014txa,Monnier:2016jlo,Monnier:2017klz,Monnier:2017oqd,Monnier:2018nfs,Monnier:2018cfa}.

As we have recalled, there are now two ways to understand the mod-12, mod-6, mod-3 conditions for $G=SU(2)$, $SU(3)$, and $G_2$.
One is from the absence of global gauge anomalies associated to non-trivial elements of $\pi_6(G)$,
and the other is from the proper consideration of the Green-Schwarz term cancelling the fermion anomaly.
The aim of this rather technical paper is to reconcile these two points of view.

\bigskip

The rest of the paper is organized as follows:
In Section \ref{sec:anom}, we carefully study all possible anomalies of theories with 0-form symmetry $G=SU(2)$, $SU(3)$, $G_2$, 
and theories with 3-form $U(1)$ symmetry.
This is done by finding the integral basis of the space of anomaly polynomials with these symmetries.
This allows us to deduce the necessary and sufficient condition when the anomalies of fermions charged under $G=SU(2)$, $SU(3)$, $G_2$ can be canceled by the anomalies of 2-form fields.
The analysis in this section does not use the homotopy groups $\pi_6(G)$ at all.

In Section \ref{sec:homotopy}, we move on to study how the derivation in Section \ref{sec:anom} is related to  the homotopy groups $\pi_6(G)$.
We do this  by carefully reformulating  
the approach of Elitzur and Nair \cite{Elitzur:1984kr}
%used to compute the global anomaly of $G$ in $6d$
in a more modern language.
The old computations of global anomalies associated to $\pi_6(G)$ can then be re-interpreted in two ways.
The first interpretation, which we give in Sec.~\ref{sec:take1}, is simply the following: 
if we assume that the perturbative anomaly is canceled by the Green-Schwarz mechanism,
the Elitzur-Nair method shows that there is a global anomaly associated to $\pi_6(G)=\bZ_k$.
But there should not be any global anomaly, since $\Omega^\text{spin}_7(BG)=0$.
This means that it is impossible to cancel the fermion anomaly by the Green-Schwarz mechanism
unless a mod-$k$ condition is satisfied.
The second interpretation, which we give in Sec.~\ref{sec:take2}, is more geometric:
we introduce a 3-form field $H$ satisfying $dH=c_2(F)$ to the bulk $7d$ spacetime,
mimicking an important half of the Green-Schwarz mechanism.
This modifies the bordism group to be considered in the classification of anomalies,
and then there actually is a global gauge anomaly associated to $\pi_6(G)$,
which defines a non-trivial element in the modified bordism group.

Before proceeding, we pause here to mention that the arguments in Section~\ref{sec:homotopy} which use $\pi_d(G)$ only give necessary conditions.
The derivation given in Section~\ref{sec:anom}, in contrast, gives conditions which are necessary and sufficient at the same time. 
In this sense, we consider the latter is better.

Furthermore, the discussions we provide in Section~\ref{sec:anom} have  already been essentially given in \cite{Monnier:2018nfs,Monnier:2018cfa}, 
albeit in a slightly different form.
Therefore, strictly speaking, our discussions in this paper do not add anything scientifically new.
That said, the authors were very much confused when they encountered the apparent contradiction that 
people discussed global gauge anomalies associated to non-trivial $\pi_6(G)$, 
while $\Omega^\text{spin}_7(BG)$ is trivial and therefore there should not be any global gauge anomalies to start with.
The authors wanted to record their understanding of how this contradiction is resolved, for future reference.

We have four appendices.
In Appendix \ref{sec:group-theoretic}, we summarize basic formulas of fermion anomalies 
and group-theoretical constants.
In Appendix \ref{sec:AHSS} and \ref{sec:Adams}, 
we compute various spin bordism groups of our interest 
using the Atiyah-Hirzebruch spectral sequence (AHSS)
and the Adams spectral sequence (Adams SS), respectively.
Finally in Appendix \ref{sec:memory}, one of the authors (Yuji Tachikawa) would like 
to share some of his recollections of his advisor, the late Professor Tohru Eguchi,
to whose memory this paper is dedicated.

\paragraph{Note added:} When this paper is almost completed, the authors learned that there is an upcoming work by Davighi and Lohitsiri \cite{DLtoAppear}, which has a large overlap with this paper.

\section{Anomalies and their cancellation}
\label{sec:anom}
In this paper, we are interested in the anomalies of fermionic $6d$ theories with $G$ symmetry, where $G=SU(2)$, $SU(3)$, and $G_2$.
To cancel them, we are also interested in the anomaly of 2-form fields, which can couple to background 4-form field strength $G$,
which is the background field for their $U(1)$ 3-form symmetry.\footnote{%
The authors apologize that the same symbol $G$ is used in three distinct ways,
for groups in general,
for the specific group $G_2$,
and for 4-form background field strengths.
Hopefully the context makes it clear which use is intended.
}
We will study when the fermion anomalies can be canceled by the anomalies of 2-form fields,
by carefully studying all possible anomalies under these symmetries.
As discussed in \eqref{AndersonDual},
the anomalies  are then characterized by the group of seven-dimensional invertible phases $\Inv^{7}_\text{spin}(X)$,
where $X=BSU(2)$, $BSU(3)$, $BG_2$ and $K(\bZ,4)$.
As $\pi_i(BE_7)=\pi_i(K(\bZ,4))$ for $i<12$, we can think of $BE_7\simeq K(\bZ,4)$ for our purposes.
This equivalence is useful for us, as we will see below, and also has been used in the past \cite{Witten:1996md,Witten:1996hc,Hsieh:2020jpj}.

Since $\Omega_7^\text{spin}(X)=0$ for all four cases, 
the anomalies are completely specified by the anomaly polynomials,
encoded in $\Hom_\bZ(\Omega_8^\text{spin}(X),\bZ)$.
Below we do not worry about purely gravitational part of the anomalies,
which means that we are going to study $\Hom_\bZ(\widetilde \Omega_8^\text{spin}(X),\bZ)$.

Over rational numbers, the anomaly polynomials are then elements of \begin{equation}
\Hom_\bZ(\widetilde \Omega_8^\text{spin}(X),\bZ)\otimes \bQ
\simeq H^8(X\times BSO;\bQ) / H^8(BSO;\bQ). \label{foo}
\end{equation}
As $H^*(BSO;\bQ)=\bQ[p_1,p_2,\ldots,]$ where $p_i \in H^{4i}(BSO;\bZ)$ are the $i$-th Pontrjagin classes of the spacetime,
Eq.~\eqref{foo} becomes \begin{equation}
%\widetilde H^8(X\times BSO;\bQ) = 
H^4(X;\bQ)\, p_1 \oplus H^8(X;\bQ).
\end{equation}
Our classifying spaces $X$ have a simplifying feature that \begin{equation}
H^4(X;\bZ)=\bZ\, c_2, \qquad H^8(X;\bZ)=\bZ\, (c_2)^2
\end{equation} where $c_2$ is a generator of $H^4(X;\bZ)$ corresponding to the instanton number,
whose  explicit forms are given in Appendix~\ref{sec:group-theoretic}.
In the rest of this paper, the Pontrjagin classes $p_i=p_i(R)$ are always for the spacetime part
and the elements $c_2=c_2(F)$ are always for the gauge part.
For $X=K(\bZ,4)$ or equivalently for $X=BE_7$, we also use the symbol $G=c_2(F)$ interchangeably.

The preceding discussions show that the anomaly polynomials (modulo purely gravitational part) are rational linear combinations \begin{equation}
\tilde\cI= a \cdot (c_2)^2 + b \cdot c_2 \wedge p_1, \label{anomrational}
\end{equation}
which has the nice quadratic form to be canceled by the Green-Schwarz mechanism.
This is, however, only a necessary condition, since the Green-Schwarz mechanism cannot cancel anomaly polynomials with arbitrary rational numbers $a$, $b$ in \eqref{anomrational}.
We need to find the basis over $\bZ$, not just over $\bQ$, of elements of $\Hom_\bZ(\Omega_8^\text{spin}(X),\bZ)$.
They are degree-8 differential forms \eqref{anomrational} which integrate to integers on any spin manifold.
We will find in Sec.~\ref{sec:KZ4}, \ref{sec:SU2}, \ref{sec:SU3}, and \ref{sec:G2}
that they are given by 
\begin{equation}
\tilde \cI = n \cdot \frac{1}{2k_G} \cdot c_2 \left(c_2+\frac{p_1}2\right) + m \cdot (c_2)^2, \qquad n,m\in \bZ
\label{preview}
\end{equation} where $k_G=12$, $6$, $3$, $1$ for $G=SU(2)$, $SU(3)$, $G_2$ and $E_7$, respectively.
We then use this result to find the anomaly cancellation condition in Sec.~\ref{sec:fermion}.
Note that $\pi_6(G)=\bZ_{k_G}$, but we \emph{do not} directly use $\pi_6(G)$ in the derivation in this section.

Before proceeding, we also note that the results analogous to \eqref{preview} pertaining to the integrality properties of the coefficients in the anomaly polynomial 
were derived in the previous literature such as \cite{Suzuki:2005vu,Kumar:2010ru} by explicitly computing the anomaly polynomials for free fermionic theories for all possible representations of $G$.
The symplectic Majorana condition was often not considered systemtically either.
As we will see below, our argument utilizes only a single representation of $G$,
thanks to the use of the bordism invariance.

\subsection{With $U(1)$  3-form symmetry}
\label{sec:KZ4}
Let us start with the case $X=K(\bZ,4)$.
%We denote the generator of $H^4(X;\bZ)$ by $G=c_2$ interchangeably.
We will find the dual bases of $\Hom_\bZ(\Omega_8^\text{spin}(X),\bZ)=\bZ\oplus \bZ$ and $\Free\Omega_8^\text{spin}(X)=\bZ\oplus \bZ$ 
by writing down two elements from each and explicitly checking that they do form a set of dual bases.\footnote{%
We note that $\Hom_{\bZ_2}(\widetilde\Omega_{10}^\text{spin}(K(\bZ,4)),\bZ_2)$ was determined in a similar manner in \cite[Sec.~3.2]{Diaconescu:2000wy}.
}

First, let us take two elements of $\Hom_\bZ(\Omega_8^\text{spin}(X),\bZ)$.
One is \begin{equation}
G \wedge G,
\end{equation}
where $G$ is a generator of $H^4(K(\bZ,4);\bZ)$.
The other element of $\Hom_\bZ(\Omega_8^\text{spin}(X),\bZ)$ we use is \begin{equation}
\frac12 \cdot G \left(G+\frac{p_1}2\right).\label{1/2}
\end{equation} There are two ways to show that it integrates to an integer on spin manifolds.
One method to demonstrate this only uses algebraic topology.
Let us first recall that the standard generator $\lambda$ of $H^4(BSpin;\bZ)=\bZ$ satisfies  $p_1=2\lambda$.
Now, for any $SO$ bundle, there is a relation
\begin{equation}
p_1 = \cP(w_2) + 2w_4 \mod 4.
\end{equation}
Here, the $2$ in front of $w_4$ is a map sending $\bZ_2=\{0,1\}$ to $\{0,2\}\subset \bZ_4$.
For spin bundles, we have $w_2=0$, and therefore $2\lambda=p_1=2w_4$,
meaning that  \begin{equation}
\lambda = w_4 \mod 2.\label{p1w4}
\end{equation} 
With this relation we can show
 \begin{equation}
	\int_{M_8} G\wedge \dfrac{p_1}{2} 
		=
		\int_{M_8} G\cup w_4
		=
		\int_{M_8} Sq^4G
		=
		\int_{M_8} G\wedge G\mod 2,
\end{equation}
where we used the fact that $w_4$ is the Wu class $\nu_4$ on a spin manifold.
Therefore the expression \eqref{1/2} integrates to an integer.

The other method is differential geometric, or uses a physics input. 
We consider the non-gravitational part $\tilde \cI_{\frac12\mathbf{56}}$ of the anomaly polynomial of a fermion in the representation $\frac12\mathbf{56}$ of $E_7$,
where $\frac12$ means that we impose a reality condition using the fact that both $\mathbf{56}$ of $E_7$ and the Weyl spinor in $6d$ with Lorentz signature are pseudo-real.
In terms of the index theory in $8d$, the same $\frac12$ uses the fact that $\mathbf{56}$ is pseudo-real and that the Weyl spinor in $8d$ with Euclidean signature are strictly real,
and therefore the index in $8d$ is a multiple of 2.
Using group-theoretical constants tabulated in Appendix~\ref{sec:group-theoretic},
the anomaly polynomial can be computed and turns out to be \begin{equation}
\tilde \cI_{\frac12\mathbf{56}}=\frac12 \cdot c_2\left(c_2+\frac{p_1}2\right).\label{56}
\end{equation} 

We next take two elements of $\Omega_8^\text{spin}(X)$, following \cite{JohnFrancis2005}.
One is the quaternionic projective space $\HP^2$.
It has a canonical $Sp(1)=SU(2)$ bundle $Q$, whose $c_2$ generates $H^4(\HP^2;\bZ)=\bZ$
such that $\int_{\HP^2} (c_2)^2=1$.
The Pontrjagin classes of the tangent bundle are $p_1=-2c_2$ and $p_2=7 (c_2)^2$.
\footnote{%
The total Pontrjagin class of $\HP^k$ is $\frac{(1-c_2)^{2(k+1)}}{(1-4c_2)}$,
see e.g.~\cite{CrowleyGoette}.
}
The other is $\CP^1\times \CP^3$ equipped with the element  $c\wedge c'\in H^4(X;\bZ)$
where $c$, $c'$ are the standard generators of $H^2(\CP^1;\bZ)$ and $H^2(\CP^3;\bZ)$, respectively.

The pairings between these elements are given by \begin{align}
\int_{(\HP^2,Q)} G\wedge G&=1,&
\int_{(\HP^2,Q)}  \frac1{2} \cdot G\left(G+\frac{p_1}2\right)&=0,\\
\int_{(\CP^1\times \CP^3,c\wedge c')} G\wedge G&=0,&
\int_{(\CP^1\times \CP^3,c\wedge c')}  \frac1{2} \cdot G\left(G+\frac{p_1}2\right)&=-1.
\end{align}
and therefore they constitute dual bases.
In other words,
the classes  $(\HP^2,Q)$ and $(\CP^1\times\CP^3,c \wedge c')$ generate $\Free \Omega_8^\text{spin}(K(\bZ,4)) = \bZ\oplus\bZ$,
while $G\wedge G$ and $ \frac1{2} \cdot G\left(G+\frac{p_1}2\right)$ generate $\Hom_\bZ(\Omega_8^\text{spin}(K(\bZ,4)),\bZ) = \bZ\oplus\bZ$.

\subsection{With $SU(2)$ symmetry}
\label{sec:SU2}
Let us next consider the case $X=BSU(2)$. 
Our approach is the same as the previous case.
We first take two elements of $\Hom_\bZ(\Omega_8^\text{spin}(X),\bZ)$.
One is $c_2\wedge c_2$ as before,
and the other is this time the non-gravitational part $\tilde \cI_{\frac12\mathbf{2}}$ of the anomaly polynomial  for the fermion in the representation $\frac12\mathbf{2}$.
Using the group theoretical data in Appendix~\ref{sec:group-theoretic}, 
we have \begin{equation}
\tilde \cI_{\frac12\mathbf{2}} = \frac1{24} \cdot c_2\left(c_2+\frac{p_1}2\right).\label{2}
\end{equation}
This can also be derived from \eqref{56} by splitting $\mathbf{56}$ as $\mathbf{2}\otimes \mathbf{12} \oplus \mathbf{1}\otimes \mathbf{32}$
under $\mathfrak{su}(2)\times \mathfrak{so}(12)\subset \mathfrak{e}_7$, 
which implies $\tilde \cI_{\frac12\mathbf{56}}=12 \cdot \tilde \cI_{\frac12\mathbf{2}}$ under $SU(2)\subset E_7$;
see e.g.~\cite[Table A.178]{Feger:2019tvk}.

We next consider two elements of $\Omega_8^\text{spin}(BSU(2))$.
One is $(\HP^2,Q)$ as before.
As the other, we take $(S^4,I)\times K3$, 
where $S^4$ is equipped with a standard instanton bundle $I$ on it, which has $\int_{S^4} c_2=1$,
while $K3$ has $\int_{K3} p_1=-48$.
Note that $S^4\simeq \HP^1$  and $I$ is its canonical $Sp(1)$ bundle.

The pairings of these elements are then 
\begin{align}
\int_{(\HP^2,Q)} c_2\wedge c_2&=1,&
\int_{(\HP^2,Q)}  \frac1{24} \cdot c_2\left(c_2+\frac{p_1}2\right)&=0,\\
\int_{(S^4,I)\times K3} c_2\wedge c_2&=0,&
\int_{(S^4,I)\times K3}  \frac1{24} \cdot c_2\left(c_2+\frac{p_1}2\right)&=-1,
\end{align}
again guaranteeing that  they form dual bases.
Equivalently, we have now shown that 
$(\HP^2,Q)$ and $(S^4,I)\times K3$ generate $\Free \Omega_8^\text{spin}(BSU(2)) = \bZ\oplus\bZ$,
and similarly $c_2\wedge c_2$ and $ \frac1{24}\cdot c_2\left(c_2+\frac{p_1}2\right)$ generate $\Hom_\bZ(\Omega_8^\text{spin}(BSU(2)),\bZ)= \bZ\oplus\bZ$.

\subsection{With $SU(3)$ symmetry}
\label{sec:SU3}
Next we consider the case $X=BSU(3)$.
Two elements of $\Hom_\bZ(\Omega_8^\text{spin}(X),\bZ)$ can be chosen as before.
One is $c_2\wedge c_2$ as always,
and the other is the non-gravitational part $\tilde \cI_{\mathbf{3}}$ of the anomaly polynomial in the representation $\mathbf{3}$.
Under $SU(3)\subset E_7$, we have $\tilde \cI_{\frac12\mathbf{56}}= 6\cdot \tilde \cI_{\mathbf{3}}$ and therefore \begin{equation}
\tilde \cI_{\mathbf{3}}=\frac{1}{12}\cdot c_2\left(c_2+\frac{p_1}2\right).
\end{equation}

A dual basis to $c_2\wedge c_2$ can be taken to be $(\HP^2,Q)$ as always.
Unfortunately, we have not found a concrete dual basis to  $\tilde \cI_{\mathbf{3}}$ in $\Omega_8^\text{spin}(X)$.
Instead we need to proceed indirectly, using the AHSS.
As computed in Appendix~\ref{sec:AHSS},
 the entries $E^2_{p,8-p}$ of the $E^2$ page for various $X$ are given by \begin{equation}
 \label{table}
\begin{array}{c|ccccccccccc}
X& E^2_{4,4} & E^2_{6,2} & E^2_{7,1} & E^2_{8,0} \\
\hline 
BSU(2) & \bZ & & & \bZ \\
BSU(3) & \bZ &  \bZ_2 &  & \bZ \\
BG_2 & \bZ & \bZ_2 & \bZ_2 & \bZ \\
K(\bZ,4) & \bZ & \bZ_2 & \bZ_2 & \bZ\oplus \bZ_3 
\end{array}
\end{equation} where the other $E^2_{p,8-p}$ are trivial.
%Under $BSU(2)\to BSU(3)\to BG_2\to K(\bZ_,4)$ the non-trivial entries in the above table are mapped isomorphically. Refs?
%Now, 
From the AHSS, it is clear that $E^2=E^\infty$ for $X=BSU(2)$, and indeed 
$\Omega^\text{spin}_8(BSU(2))=\bZ\oplus \bZ$.
We also know $\Free\Omega^\text{spin}_8(K(\bZ,4))=\bZ\oplus \bZ$
from the explicit analysis in Sec.~\ref{sec:KZ4}.
Comparing the bases in Sec.~\ref{sec:KZ4} and \ref{sec:SU2},
we find that under $BSU(2)\to K(\bZ,4)$, the image of $\Omega^\text{spin}_8(BSU(2))$
is $12\bZ\oplus \bZ\subset \bZ\oplus \bZ$.
This means that the extension problem in the AHSS for $K(\bZ,4)$ is solved as follows:
we have \begin{equation}
0\to \underbrace{E^{2}_{4,4} \oplus \Free E^2_{8,0}}_{=\bZ\oplus\bZ} \to 
\underbrace{\Omega^\text{spin}_8(K(\bZ,4))}_{=\bZ\oplus\bZ} \to \bZ_{12} \to 0
\label{extension-problem}
\end{equation}
where $\bZ_{12}$ is composed of 
$E^{2}_{6,2}=\bZ_2$, $E^{2}_{7,1}=\bZ_2$, and $\Tors E^{2}_{8,0}=\bZ_3$ in the last row of \eqref{table}.

Since $BSU(3)$ is sandwiched as in  $BSU(2)\to BSU(3) \to K(\bZ,4)$,
we conclude that $\Omega^\text{spin}_8(BSU(3)) =\bZ\oplus \bZ$
whose image under $BSU(3)\to K(\bZ,4)$ is $6\bZ\oplus \bZ \subset \bZ\oplus\bZ$.
This abstract analysis provides a dual basis element to $\tilde \cI_{\mathbf{3}}= \frac16 \cdot\tilde \cI_{\frac12\mathbf{56}}$.

\subsection{With $G_2$ symmetry}
\label{sec:G2}
The $G_2$ case is completely analogous to the $SU(3)$ case.
One basis of $\Hom_\bZ(\Omega^\text{spin}_8(BG_2),\bZ)=\bZ\oplus\bZ$
is given by $c_2\wedge c_2$ as always and the other is given by \begin{equation}
\tilde \cI_{\mathbf{7}}=\frac1{6} \cdot c_2\left(c_2+\frac{p_1}2\right),
\end{equation} which satisfies $\tilde \cI_{\mathbf{7}}=\frac13 \cdot \tilde \cI_{\frac12\mathbf{56}}$ under $G_2\subset E_7$.
The dual basis to $c_2\wedge c_2$ is given by $(\HP^2,Q)$.
The existence of the dual basis to $\tilde \cI_{\mathbf{7}}$ can be argued exactly as before,
and the image of $\Omega^\text{spin}_8(BG_2)=\bZ\oplus\bZ$
under $BG_2\to K(\bZ,4)$  is $3\bZ\oplus \bZ \subset \bZ\oplus\bZ$.

\subsection{Pure gravitational part}
Before proceeding, here we record the dual bases of
$\Hom_\bZ(\Omega^\text{spin}_8(pt),\bZ)=\bZ\oplus \bZ$
and $\Omega^\text{spin}_8(pt)=\bZ\oplus \bZ$.
Two generators of the latter were found e.g.~in \cite{MilnorSpin},
which are $\HP^2$ and $L_8$, where $4L_8$ is spin bordant to $K3\times K3$.
They have Pontrjagin numbers 
$p_1^2(\mathbb{HP}^2)=4$,
$p_2(\mathbb{HP}^2)=7$
and 
$p_1^2(L_8)=1152$, 
$p_2(L_8)=576$,
respectively.
As for the generators of the former, we can take the anomaly polynomials $\cI_\text{fermion}$ and $\cI_\text{gravitino}$ of the $6d$ fermion and the gravitino,
which can be found in any textbook: \begin{equation}
\cI_\text{fermion}=\frac{7\,p_1^2-4\, p_2}{5760},\qquad
\cI_\text{gravitino}=\frac{275\,p_1^2-980\, p_2}{5760}.
\label{ferano}
\end{equation}
They have the pairing \begin{align}
\int_{\HP^2} \cI_\text{fermion} &=0, &
\int_{\HP^2} \cI_\text{gravitino} &=-1, \\
\int_{L_8} \cI_\text{fermion} &=1, &
\int_{L_8} \cI_\text{gravitino} &=-43
\end{align} and form a pair of dual bases.

\subsection{Anomalies of self-dual 2-forms}
\label{sec:self-dual-2form}
Now that we have discussed the general structures of anomalies with symmetries of our interest,
we would like to study their cancellation.
For this, we need to examine the anomalies of self-dual form fields in more detail.

Let us start by recalling the naive analysis often found in the older literature. 
The  one-loop anomaly polynomial of a self-dual tensor can be found e.g.~in \cite{AlvarezGaume:1983ig}.
In general dimensions, it is given by $\pm L/8$, where $L$ is the Hirzebruch genus.
In six dimensions this gives\footnote{%
Our sign choice here is for $-L/8$ while the fermion anomaly in \eqref{ferano} corresponds to $+\hat A$.
They are the anomaly polynomials for the tensor field in the tensor multiplet and for the fermion in the hypermultiplet in \Nequals{(1,0)} supersymmetry in six dimensions.
} \begin{equation}
\cI_\text{tensor,one-loop}=\frac{16\,p_1^2-112\,p_2}{5760}.
\end{equation} 
Furthermore, at the level of differential forms,
a self-dual tensor $B$ can be coupled to a background 4-form $\tilde G$ via $dH=\tilde G$,
which contributes to the anomaly polynomial by $\frac{1}{2} \tilde G\wedge \tilde G$.
Then the total anomaly is \begin{equation}
\cI_\text{tensor} =\cI_\text{tensor,one-loop} + \frac12 \tilde G\wedge \tilde G.
\end{equation}
Note that this \emph{does not} integrate to integers on $\HP^2$ and $L_8$ when we naively set $\tilde G=0$.
Rather, we need to set \begin{equation}
\tilde G=\frac{p_1}4
\label{p1/4}
\end{equation}
 for which we find \begin{equation}
\cI_\text{tensor}=\frac{196\,p_1^2-112\,p_2}{5760}=28\cdot \cI_\text{fermion},
\end{equation}
which \emph{does} integrate to integers on $\HP^2$ and $L_8$.
We cannot, however, promote $\tilde G$ to be a 3-form field strength, since $\frac{p_1}{4}$ is not integrally quantized in general, although $\frac{p_1}{2}$ is, as we saw above.

A proper formulation was first proposed by Belov and Moore in \cite{Belov:2006jd} using the results of Hopkins and Singer \cite{Hopkins:2002rd},
and it requires the use of Wu structure in general $(4k+2)$ dimensions.
This formulation  was then developed in detail in a series of papers by Monnier and collaborators \cite{Monnier:2010ww,Monnier:2011mv,Monnier:2011rk,Monnier:2012xd,Monnier:2013kna,Monnier:2013rpa,Monnier:2014txa,Monnier:2016jlo,Monnier:2017klz,Monnier:2017oqd,Monnier:2018nfs,Monnier:2018cfa}. 
For our case of interest in six spacetime dimensions, a spin structure induces a canonical Wu structure,
and self-dual form fields can be naturally realized on the boundary of a natural class of $7d$ topological field theories \cite{Hsieh:2020jpj};
an in-depth discussion can also be found in \cite{Gukov:2020btk}.

In the case of a single self-dual form field of level 1, the bulk theory has the action \begin{equation}
S=-\dint_{M_7}\left(
		\dfrac{1}{2}\cdot c \left(dc
		+
		\dfrac{p_1}{2}
		\right)
		+
		c \wedge G
	\right)
	\label{7daction}
\end{equation}
where $c$ is a 3-form gauge field to be path-integrated and $G$ is a background 4-form field strength.
It should more properly be written using its $8d$ extension as 
\begin{equation}
S=-\dint_{M_8}\left(
		\dfrac{1}{2}\cdot g \left(g
		+
		\dfrac{p_1}{2}
		\right)
		+
		g \wedge G
	\right)
	\label{8daction}
\end{equation} where $M_8$ is a spin manifold such that $\partial M_8=M_7$,
and $g=dc$ is the field strength of $c$;
we extend $g$ and $G$ to the entire $M_8$.
Such extension is guaranteed to exist since $\Omega^{\text{spin}}_7(K(\bZ^{\oplus n},4))=0$,
and the action does not depend on the choice of the extension 
since we know that  $\frac12\cdot g(g+\frac{p_1}{2})$ and $g\wedge G$ integrate to integers on spin manifolds,
as we saw in Sec.~\ref{sec:KZ4}.

At the level of differential forms, one is tempted to rewrite \eqref{7daction} as \begin{equation}
S=-\dint_{M_7}\left(
		\dfrac{1}{2}\cdot c\wedge dc
		+c \wedge \tilde G
	\right)
\end{equation}
where \begin{equation}
\tilde G=\frac{p_1}4 + G.
\end{equation} 
In this sense, a single self-dual 2-form is forced to couple to $\frac{p_1}4$ from its consistency,
explaining the observation \eqref{p1/4} above.
In this paper, we stick to use the unshifted $G$ which is integrally quantized.

In \eqref{8daction}, we can introduce a new variable $\hat g= g+G$ and rewrite it as \begin{equation}
S=-\dint_{M_8} \dfrac{1}{2}\cdot \hat g \left(\hat g
		+
		\dfrac{p_1}{2}
		\right)
+\dint_{M_8} \dfrac{1}{2}\cdot G \left( G
		+
		\dfrac{p_1}{2}
		\right) .
\end{equation}
As $\hat g$ is path-integrated, the first term produces the pure gravitational anomaly of a single self-dual tensor field `properly coupled to $\frac14p_1$',
and the second term is the residual dependence on $G$,
which we saw in Sec.~\ref{sec:KZ4} to be a generator of $\Hom_\bZ(\Omega^\text{spin}_8(K(\bZ,4)),\bZ)$.

The other generator was $G\wedge G$, which can be realized as the standard anomaly of a non-chiral 2-form field coupled to $G$ both electrically and magnetically.
Combining these two pieces of information,
we find that every possible anomaly of $\Hom_\bZ(\Omega^\text{spin}_8(K(\bZ,4)),\bZ)$
can be realized by a system of 2-form fields in $6d$.
More generally,
if there are $n$ self-dual 2-form fields, the $7d$ bulk TQFT has the form
\begin{equation}
	S=
	-\dint_{M_8}\left(
		\dfrac{1}{2}\left(K_{ij}\cdot g^i \wedge g^j
		+
		a_i\cdot g^i\wedge \dfrac{p_1}{2}\right)
		+
		g^i \wedge Y_i.
	\right)
\end{equation}
Here, $i$ runs from $1$ to $n$,
$c^i$ are the 3-form fields to be integrated over, 
$g^i=dc^i$ are their curvatures,
$Y_i$ are the background fields of $U(1)$ 3-form symmetry, 
and $K_{ij}$ and $a_i$ are integer coefficients.
For this expression to be well defined,
we need to require\footnote{%
This relation can be more invariantly stated as follows.
We introduce a lattice $\Lambda=\bZ^n$ whose pairing is given by $K_{ij}$.
Then the vector $(a_i)\in \Lambda$ is a characteristic vector, i.e~$\braket{x,x}\equiv \braket{a,x}$ modulo two.
That the coefficients $a_i$ multiplying $g^i\wedge \frac{p_1}2$ in the anomaly polynomial of the fermionic part of a $6d$ theory form a characteristic vector was long conjectured, 
e.g.~in \cite{Monnier:2017oqd} and finally proved in \cite{Monnier:2018nfs,Monnier:2018cfa}. 
We also note that an analogous constraint in the case of $3d$ Chern-Simons theory coupled to the $\text{spin}^c$ connection was first noted in \cite[Sec.~7]{Belov:2005ze} 
which was later utilized to a great effect in \cite[Sec.~2.3]{Seiberg:2016rsg}.
} \begin{equation}
K_{ii}=a_i \mod 2\qquad\text{(not a sum over $i$)}.
\end{equation} 
To see this, we first note
the expression $\frac12\sum_{i\neq j} K_{ij}\cdot g^i \wedge g^j = \sum_{i<j} K_{ij}\cdot g^i \wedge g^j$ clearly integrates to an integer.
Then, we can separately analyze each $i$.
We finally recall that the generators of $\Hom_\bZ(\Omega^\text{spin}_8(K(\bZ,4)),\bZ)$
are $G\wedge G$ and $\frac12\cdot G(G+\frac{p_1}2)$, and we are done.

If $\det K\neq 1$, this theory is not invertible but rather a  topological field theory with a multi-dimensional Hilbert space.
Correspondingly, the boundary $6d$ theory is a \emph{relative} theory in the sense of \cite{Freed:2012bs},
which does not have a partition function but rather a partition \emph{vector}.
The system then depends on the background fields $Y_i$ in a complicated manner,
and cannot be directly used to cancel the fermion anomaly.
However, suppose that $Y_i$ has the form $Y_i = K_{ij} \cdot G^j+\hat Y_i$. 
In this case, shifting $g^i$ to new fields $\widetilde g^i = g^i + G^i$, one can factor out as
\begin{multline}
	S=-\dint_{M_8}\left(
		\dfrac{1}{2}\left(K_{ij}\cdot \widetilde g^i \wedge \widetilde g^j
		+
		a_i \cdot \widetilde g^i\wedge \dfrac{p_1}{2}\right)
		+
		\widetilde g^i \wedge\hat Y_i
	\right)\\
	+\dint_{M_8}\left(
		\dfrac{1}{2}\left(K_{ij}\cdot G^i\wedge G^j
		+
		a_i \cdot G^i\wedge \dfrac{p_1}{2}\right)
		+
		G^i\wedge \hat Y_i 
	\right)
\end{multline}
where the second line is invertible since it only depends on the background field $G^i$.
It is this invertible part which is used to cancel the fermion anomaly, 
and the remaining non-invertible part is irrelevant for our purposes.
Setting $G^i=n^i \cdot G$ for a single $G$ and integers $n^i$,
we again see that we can produce an arbitrary element of $\Hom_\bZ(\Omega^\text{spin}_8(K(\bZ,4)),\bZ)$ as the anomaly of  a system of self-dual 2-form fields.
In the examples in M-theory and F-theory treated in \cite{Ohmori:2014kda}, 
the cancellation was indeed done in this manner, where $G^i$ were set to $c_2$ of dynamical gauge fields and $\hat Y_i$ were set to $c_2$ of the R-symmetry background.

%\newpage

\subsection{Cancellation}
\label{sec:fermion}

After all these preparations, the analysis of the cancellation is very simple.
Consider a system of fermions charged under $G=SU(2)$, $SU(3)$, or $G_2$.
From the results of Sec.~\ref{sec:SU2}, \ref{sec:SU3} and \ref{sec:G2},
its anomaly is necessarily of the form \begin{equation}
\tilde \cI_\text{fermion} = n \cdot \frac{1}{2k_G} \cdot c_2\left(c_2+\frac{p_1}2\right) + m \cdot c_2\wedge c_2
\label{fermion-summary}
\end{equation}
where $n$ and $m$ are integers and 
$k_G=12$, $6$, $3$ for $G=SU(2)$, $SU(3)$, and $G_2$, respectively.

Now, take a number of self-dual 2-form fields and use $c_2$ as their background field $G$.
As we saw in Sec.~\ref{sec:self-dual-2form}, the form of anomalies which can be produced by a system of 2-form fields has the structure \begin{equation}
\tilde \cI_\text{tensor}= \hat n \cdot \frac{1}{2} \cdot G\left(G+\frac{p_1}2\right) + \hat m \cdot G \wedge G,
\end{equation}
where $\hat n$ and $\hat m$ are again integers.
This means that the necessary and sufficient condition for the anomaly of fermions to be cancelable by the anomaly of 2-form fields is \begin{equation}
n \equiv 0 \mod k_G.\label{condition}
\end{equation}
Here, $\pi_6(G)=\bZ_{k_G}$ for our $G$,
but we did not use this in the derivation presented in this section.

We note that the coefficient $n$ controls the fractional part of the coefficient of $c_2\wedge c_2$, which comes from the $\tr F^4$ term in the fermion anomaly.
For $G=SU(2)$, $SU(3)$, and $G_2$, we have \begin{equation}
\mathrm{tr}_{\text{rep.}}F^4
=
		\gamma_{\text{rep.}}\cdot \big(\mathrm{tr}_{\text{fund.}}F^2\big)^2,
\end{equation}
and we saw in Sec.~\ref{sec:SU2}, \ref{sec:SU3}, and \ref{sec:G2}
that a fermion in the representation $\frac12\mathbf{2}$, $\mathbf{3}$, and $\mathbf{7}$ 
corresponds to $n=1$ in \eqref{fermion-summary}.
Therefore, the condition \eqref{condition} can be expanded more concretely into
%the coefficients of the first term of \eqref{anomaly-fermion} must be integral, i.e.
\begin{equation}
	\label{perturbative-anomaly-cancellation}
	\renewcommand{\arraystretch}{1.3}
	\begin{array}{cl}
		SU(2):
		& \displaystyle\sum_{i^+} \frac{\gamma_{\text{rep.$(i^+)$}}}{\gamma_{\frac12\mathbf{2}}} - \sum_{i^-} \frac{\gamma_{\text{rep.$(i^-)$}}}{{\gamma_{\frac12\mathbf{2}}}} = 0 \mod 12,\\
		SU(3): 
		& \displaystyle\sum_{i^+} \frac{\gamma_{\text{rep.$(i^+)$}}}{\gamma_{\mathbf3}} - \sum_{i^-} \frac{\gamma_{\text{rep.$(i^-)$}}}{\gamma_{\mathbf{3}}} = 0 \mod 6,\\
		G_2:
		& \displaystyle\sum_{i^+} \frac{\gamma_{\text{rep.$(i^+)$}}}{\gamma_{\mathbf7}} - \sum_{i^-} \frac{\gamma_{\text{rep.$(i^-)$}}}{\gamma_{\mathbf7}} = 0 \mod 3,\\
	\end{array}
\end{equation}
where $\gamma_{\frac12\mathbf{2}}=1/4$, $\gamma_{\mathbf{3}}=1/2$, and $\gamma_{\mathbf{7}}=1/4$, and
the summation runs over all fermions in the theory where $+/-$ represents their chiralities.
Some values of $\gamma_\text{rep.}$ are collected in  Appendix~\ref{sec:group-theoretic}.

\subsection{Examples: $\cN=(1,0)$ supersymmetric cases}
\label{sec:example-SUSY}
Let us provide some concrete examples of the anomaly cancellation.
In $6d$ $\cN=(1,0)$ QFT, fermions charged under the gauge group are gaugini in the adjoint representation and hyperini in the matter representation.
They are in opposite chiralities. 
Let us restrict the representations of hyperini and denote their numbers as follows:
\begin{equation}
	\begin{array}{cccc}
		n_2 & : & SU(2) & \text{fundamental},\\
		n_3 & : & SU(3) & \text{fundamental},\\
		n_6 & : & & \text{symmetric},\\
		n_7 & : & G_2 & \text{fundamental}.\\
	\end{array}
\end{equation}
Then, the conditions for the perturbative anomaly cancellation \eqref{perturbative-anomaly-cancellation} are
\begin{equation}
	\label{example-SUSY}
	\begin{array}{cclll}
		SU(2) & : 
		& 4\gamma_{\text{adj.}} - 4\gamma_{\text{fund.}} \cdot n_2
		& = 32 - 2\cdot n_2 
		& = 0 \mod 12,\\
		SU(3) & :
		& 2\gamma_{\text{adj.}} - 2\gamma_{\text{fund.}} \cdot n_3 -  2\gamma_{\text{sym.}}\cdot n_6
		& = 18 - n_3 - 17\cdot n_6 
		& = 0 \mod 6,\\
		G_2 & :
		& 4\gamma_{\text{adj.}} - 4\gamma_{\text{fund.}} \cdot n_7
		& = 10 - n_7
		& = 0 \mod 3.\\
	\end{array}
\end{equation}
which exactly coincide with the conditions for absence of global anomalies appeared in \cite{Bershadsky:1997sb}.

\section{Relation to $\pi_6(G)$}
\label{sec:homotopy}
In the last section, we deduced the condition when the anomaly of a fermion system with $G$ symmetry can be canceled by the anomaly of 2-form fields, for $G=SU(2)$, $SU(3)$, and $G_2$.
Our exposition used a careful determination of integral generators of the group of anomalies.
In the end, we found a mod-12, mod-6, and mod-3 condition, respectively.

Traditionally, the same anomaly cancellation condition was related to the anomaly by a global gauge transformation associated to $\pi_6(G)=\bZ_{12}$, $\bZ_6$, and $\bZ_3$ 
for each $G$.
However, our derivation in the last section did not use $\pi_6(G)$ at all.
In this section, we would like to clarify the relation between these two approaches.

\subsection{The Elitzur-Nair method}
\label{sec:EN}
Let us first remind ourselves the method to determine the global gauge anomaly of a system under $G$ symmetry, assuming the cancellation of perturbative gauge anomalies,
by rewriting it as a perturbative anomaly under a larger group $\tilde G$ whose global gauge anomaly is absent.
This method is originally due to \cite{Elitzur:1984kr}.
In the context of the modern bordism approach to the anomalies, it was re-discovered 
in \cite{Davighi:2020bvi} and further developed more extensively in \cite{Davighi:2020uab}.
The basic idea is as follows.

The global $G$ gauge anomaly may arise due to gauge transformations $g:S^d\to G$ belonging to non-trivial elements of $\pi_{d}(G)=\bZ_k$,
which cannot be continuously deformed to the trivial transformation.
The first step is then to embed the gauge group $G$ into a larger group $\tilde G$ with $\pi_d(\tilde G)=0$.
This means that $g$ can be trivialized within $\tilde G$.
Further, assume that our $G$-symmetric system $Q$ is obtained by restricting the symmetry of a $\tilde G$-symmetric system $\tilde Q$.
Using this, one can describe the global $G$ gauge anomaly of $Q$ in terms of perturbative $\tilde G$ gauge anomaly of $\tilde Q$.

Let us implement this idea, following \cite{Elitzur:1984kr}.
We first note that the fibration
\begin{equation}
	G
	\ \textarrow{$\iota$}\ 
	\tilde G
	\ \textarrow{$p$}\ 
	\tilde G/G 
	\label{fib}
\end{equation}
induces a homotopy exact sequence,
which we assume to have the form
\begin{equation}
	\cdots
	\ \textarrow{$\iota_\ast$}\ 
	\underbrace{\pi_{d+1}(\tilde G)}_{=\bZ}
	\ \textarrow{$p_\ast$}\ 
	\underbrace{\pi_{d+1}(\tilde G/G)}_{=\bZ}
	\ \textarrow{$\partial$}\ 
	\underbrace{\pi_d(G)}_{=\bZ_k}
	\ \textarrow{$\iota_\ast$}\ 
	\underbrace{\pi_d(\tilde G)}_{=0}
	\ \textarrow{$p_\ast$}\ 
	\cdots
\end{equation}
so that the generator of $\pi_{d+1}(\tilde G)$ is mapped to a $k$-th power of the generator of $\bZ= \pi_{d+1}(\tilde G/G)$. 
Now, to compute the anomaly under $g:S^d\to G$,
we consider a $G$ bundle  on $S^{d+1}$ obtained by gluing two trivial bundles on two hemispheres at the equator $S^d\subset S^{d+1}$.
The resulting $G$ bundle on $S^{d+1}$ is classified by an element $\pi_{d+1}(BG)$, 
which is naturally isomorphic to $\pi_d(G)$.
We denote the resulting $G$ bundle by $P_{[g]}$ where $[g]\in \pi_d(G)\simeq \pi_{d+1}(BG)$.
In the modern understanding, the anomaly is the partition function $Z_{A(Q)}[S^{d+1},P_{[g]}] $ of the $(d+1)$-dimensional anomaly theory $A(Q)$ evaluated on $(S^{d+1},P_{[g]})$.

From the assumption that $\pi_{d}(\tilde G)=\pi_{d+1}(B\tilde G)=0$,
we can extend this $G$ bundle $P_{[g]}$ on $S^{d+1}$ to 
a $\tilde G$ bundle on $D^{d+2}$.
Such an extension is classified by an element  $[f]\in \pi_{d+1}(\tilde G/G)$
such that $\partial[f]=[g]$.
Denoting the corresponding $\tilde G$ bundle on $D^{d+2}$ by $\tilde P_{[f]}$,
the anomaly is $\exp(2\pi i J([f]))$ where \begin{equation}
J([f])=\int_{(D^{d+2},\tilde P_{[f]})} \mathcal{I}_{\tilde Q} \label{fQ}
\end{equation}
where $\mathcal{I}_{\tilde Q}$ is the anomaly polynomial of the $\tilde G$-symmetric theory $\tilde Q$.
One can check that the expression \eqref{fQ} gives a well-defined homomorphism \begin{equation}
J: \pi_{d+1}(\tilde G/G)\to \bR,
\end{equation} 
since $\mathcal{I}_{\tilde Q}$ is a closed form and its restriction to $G$ bundles is zero by assumption.
Let us now take a generator $[f_0]\in \pi_{d+1}(\tilde G/G) = \bZ$,
which maps to a generator $[g_0]=\partial[f_0] \in \pi_{d+1}(BG) = \bZ_k$.
Then the global anomaly we are after is 
\begin{equation}
Z_{A(Q)}[S^{d+1},P_{[g_0]}] = \exp\Big(2\pi i J([f_0])\Big).
\end{equation}
To compute it, note that $k[f_0] \in \pi_{d+1}(\tilde G/G)$ is the image of the generator 
$[\tilde g_0] \in \pi_{d+1}(\tilde G)\simeq\pi_{d+2}(B\tilde G)$.
In this case, we can attach  
a trivial $\tilde G$ bundle over another copy of $D^{d+2}$  
to the bundle $(D^{d+2},\tilde P_{k[f_0]})$ discussed above 
along the common boundary $S^{d+1}$ via gauge transformation in $[\tilde g_0]$,
to form a $\tilde G$ bundle $\hat P_{[\tilde g_0]}$ on $S^{d+2}$.
Then \begin{equation}
J(k[f_0])= \int_{(D^{d+2},\tilde P_{k[f_0]})} \mathcal{I}_{\tilde Q} 
= \int_{(S^{d+2},\hat P_{[\tilde g_0]})} \mathcal{I}_{\tilde Q}.
\end{equation} 
This last expression is often computable,
resulting in the final formula that 
the anomalous phase associated to the gauge transformation in $[g_0]\in \pi_d(G)$ is given by \begin{equation}
Z_{A(Q)}[S^{d+1},P_{[g_0]}] 
=\exp\left(
	\frac{2\pi i}k \int_{(S^{d+2},\hat P_{[\tilde g_0]})} \mathcal{I}_{\tilde Q}
\right).
\end{equation}
This derivation was originally devised by Elitzur and Nair \cite{Elitzur:1984kr} for the $4d$ analysis, 
where Witten's original anomaly in the $G=SU(2)$ case was derived by  embedding $G$ to $\tilde G=SU(3)$.
It was also re-discovered in the context of our recent progress in the understanding of anomalies in the bordism point of view by \cite{Davighi:2020bvi}, 
who embedded $SU(2)$ to $U(2)$ instead.%, apparently without knowing the original work by Elitzur and Nair.

Let us quickly see how Witten's anomaly is computed in this framework.
The sequence is 
\begin{equation}
	\cdots
	\ \textarrow{$\iota_\ast$}\ 
	\underbrace{\pi_{5}(SU(3))}_{=\bZ}
	\ \textarrow{$p_\ast$}\ 
	\underbrace{\pi_{5}(SU(3)/SU(2))}_{=\bZ}
	\ \textarrow{$\partial$}\ 
	\underbrace{\pi_4(SU(2))}_{=\bZ_2}
	\ \textarrow{$\iota_\ast$}\ 
	\underbrace{\pi_4(SU(3))}_{=0}
	\ \textarrow{$p_\ast$}\ 
	\cdots,
\end{equation}
where $SU(3)/SU(2)\simeq S^5$.
We take the doublet in $SU(2)$ minus two uncharged Weyl spinors as the theory $Q$,
and the triplet in $SU(3)$ minus three uncharged Weyl spinors as the theory $\tilde Q$.
By restricting the symmetry from $SU(3)$ to $SU(2)$, the theory $\tilde Q$ reduces to $Q$,
since an uncharged pair of Weyl spinors of different chirality can be given a mass 
and therefore does not contribute to the anomaly.
The anomaly polynomial of $\tilde Q$ is simply $\frac1{3!}\tr \left(\frac{F}{2\pi}\right)^3$,
which is known to integrate to $1$ against the generator $[\tilde g_0]$ of $\pi_5(SU(3))=\pi_6(BSU(3))$,
meaning that $J(2[f_0])=1$.
Therefore, the anomalous phase under the generator $[g_0]\in \pi_4(SU(2))$ is 
$\exp(2\pi i J([f_0]))=\exp(2\pi i \cdot \frac{1}{2})=-1$.
In this way, we see that the anomaly theory of a $4d$ fermion in the doublet of $SU(2)$ detects the generator of $\Omega^\text{spin}_5(BSU(2))=\bZ_2$.

\subsection{$6d$  analysis, take 1}
\label{sec:take1}

The analysis of Elitzur and Nair was applied and extended to the $6d$ case in \cite{Tosa:1989qm, Bershadsky:1997sb,Suzuki:2005vu}.
However, the analyses there were not quite satisfactory from a more modern point of view,
since the contribution from the self-dual tensor field was not properly taken care of.
The first objection to their analyses is the following:
we know that $\pi_6(SU(2))=\bZ_{12}\to \Omega^\text{spin}_7(BSU(2))=0$ is a zero map. 
How can there be a global anomaly?
The second objection is the following:
we know from Sec.~\ref{sec:fermion} that we cannot cancel the perturbative anomaly of a single fermion in $\frac12\mathbf{2}$ of $SU(2)$ by that of self-dual tensor fields.
How can we apply the Elitzur-Nair method?
We can combine these two objections into a consistent modern reinterpretation of the old analyses, as follows.

Suppose that the perturbative anomaly of $n$ fermions in $\frac12\mathbf{2}$ of $SU(2)$ can be canceled by a combination of 2-form fields, by using $c_2$ of the $SU(2)$ bundle as the background of the $U(1)$ 3-form symmetry of the 2-form fields.
We also assume that the perturbative gravitational anomaly is canceled by adding some uncharged fields,
and take the combined theory as $Q$.
Since the perturbative anomaly vanishes, it determines a bordism invariant \begin{equation}
\Omega^\text{spin}_7(BSU(2)) \to U(1),
\end{equation} 
but as $\Omega^\text{spin}_7(BSU(2))=0$, it is trivial.
In particular, the anomalous phase associated to $(S^7,P_{[g_0]})$ for the generator $[g_0] \in \pi_7(BSU(2))=\bZ_{12}$ should also be trivial, i.e. \begin{equation}
Z_{A(Q)}[S^{7},P_{[g_0]}] = 1.\label{...}
\end{equation}

We now compute this anomalous phase by the Elitzur-Nair method.
Taking $G=SU(2)=Sp(1)$ and $\tilde G=Sp(2)$, the sequence is
\begin{equation}
	\cdots
	\ \textarrow{$\iota_\ast$}\ 
	\underbrace{\pi_{7}(Sp(2))}_{=\bZ}
	\ \textarrow{$p_\ast$}\ 
	\underbrace{\pi_{7}(Sp(2)/Sp(1))}_{=\bZ}
	\ \textarrow{$\partial$}\ 
	\underbrace{\pi_6(Sp(1))}_{=\bZ_{12}}
	\ \textarrow{$\iota_\ast$}\ 
	\underbrace{\pi_6(Sp(2))}_{=0}
	\ \textarrow{$p_\ast$}\ 
	\cdots,
	\label{Sp1seq}
\end{equation}
where $Sp(2)/Sp(1)\simeq S^7$.
As the theory $\tilde Q$, we take $n$ copies of $\frac12\mathbf{4}$ of $Sp(2)$, 
coupled to the same combination of 2-form fields, by using $c_2$ of the $Sp(2)$ bundle as the background of the $U(1)$ 3-form symmetry.
Note that our presentation of the Elitzur-Nair method in Sec.~\ref{sec:EN} applies to this case.
The anomaly polynomial $\cI_{\tilde Q}$ of $\tilde Q$ is $n$ times $\frac12\cdot\frac{1}{4!} \tr \left(\frac{F}{2\pi}\right)^4$ modulo multi-trace terms.
Now, it is  known that the expression $\frac12\cdot\frac{1}{4!} \tr \left(\frac{F}{2\pi}\right)^4$  integrates to $1$ against the generator $[\tilde g_0]$ of $\pi_7(Sp(2))=\pi_8(BSp(2))=\bZ$\cite[Appendix B.2]{Suzuki:2005vu}.
This means that for $\partial[\tilde g_0]=12[f_0]$ we have \begin{equation}
J(12[f_0])
= \int_{(S^{d+2},\hat P_{[\tilde g_0]})} \cI_{\tilde Q}
= n \int_{(S^{d+2},\hat P_{[\tilde g_0]})} \dfrac{1}{2}\cdot \frac{1}{4!} \tr \left(\frac{F}{2\pi}\right)^4 = n.
\end{equation}
Therefore the anomalous phase associated to $(S^7,P_{[g_0]})$ 
is \begin{equation}
Z_{A(Q)}[S^{7},P_{[g_0]}] = \exp\left(n \cdot \frac{2\pi i}{12}\right)\label{,,,}.
\end{equation}

For the results \eqref{...} and \eqref{,,,} to be consistent, 
we conclude that $n$ needs to be divisible by $12$.
The analysis can be extended to arbitrary fermions charged under $SU(2)$,
and we can re-derive  \eqref{perturbative-anomaly-cancellation} as a necessary condition.
The argument in this subsection, however, does not say that they are also sufficient.
Differently put,  the method in this section does not directly show that the fermion anomaly can actually be canceled, without the analysis of Sec.~\ref{sec:fermion}.

The analysis for $G=SU(3)$ and $G_2$ can be rephrased in a completely similar manner 
by choosing $\tilde G$ to be $SU(4)$ and $SU(7)$ respectively,
so will not be repeated.

\subsection{$6d$ analysis, take 2}
\label{sec:take2}
The argument in the previous subsection can also be phrased in the following manner.
Let us start by assuming that the perturbative anomaly of a single fermion in $\frac12\mathbf{2}$ can be canceled.
Then, we can apply the Elitzur-Nair argument to compute the anomaly under the generator $[g_0]\in \pi_6(SU(2))=\bZ_{12}$,
which is the partition function of the anomaly theory on $(S^7,P_{[g_0]})$.
The value turns out to be $\exp(2\pi i/12)$.
The configuration $(S^7,P_{[g_0]})$ is however null-bordant in $\Omega^\text{spin}_7(BSU(2))$,
and therefore the partition function should be 1.
This leads to a contradiction,
meaning that the initial assumption is incorrect.
We conclude that the perturbative anomaly cannot be canceled for a single fermion in $\frac12\mathbf{2}$.

This argument still leaves us wondering whether it is possible to have a setup where the anomaly associated to $[g_0]$ is actually $\exp(2\pi i/12)$.
It turns out that it can be achieved by considering a slightly different bordism group as follows. 
This is essentially the computation done in \cite{Tosa:1989qm,Bershadsky:1997sb}.

So far, we have considered the manifolds in question to be equipped with spin structure and a $G$ gauge field.
Let us equip the manifolds with a classical 3-form field $H$, 
which we consider to be  a (modified) field strength of the 2-form field $B$, satisfying 
\begin{equation}
	\label{dH=c2}
	dH = c_2.
\end{equation}
This relation makes 
the part of the anomaly polynomial divisible by $c_2$ cohomologically trivial, and effectively cancels it.
Let us see this more concretely.
Take $G=SU(2)$ and consider fermions in $\frac12\mathbf{2}$.
The partition function of its anomaly theory $A(Q)$ is given by the associated eta invariant $\eta_{\frac12\mathbf{2}}$,
whose variation is controlled by the anomaly polynomial \begin{equation}
\frac{1}{24}\cdot c_2\left(c_2+\frac{p_1}2\right).
\end{equation}
With \eqref{dH=c2} we can add a local counterterm in $7d$, and consider the anomaly theory to be \begin{equation}
Z_{A(Q)'} = \exp\left(2\pi i \left(\eta_{\frac12\mathbf{2}} - \int_{M_7} \frac{1}{24} \cdot H \left(c_2+\frac{p_1}2\right)\right)\right). \label{modification}
\end{equation}
This combination is clearly constant under infinitesimal variations of the background fields,
and gives a invariant of the bordism group equipped with spin structure, a $G$ gauge field, and a 3-form field satisfying \eqref{dH=c2}.

The $G$-field configuration $P_{[g_0]}$ on $S^7$ has a cohomologically-trivial $c_2$, since $H^4(S^7)$ vanishes. 
Therefore we can solve \eqref{dH=c2} to find $H$ on $S^7$.
The $\tilde G$-field configuration $\tilde P_{[f_0]}$ on $D^8$ can also be augmented with a solution to \eqref{dH=c2}, since $H^4(D^8)$ is again trivial. 
Then the Elitzur-Nair argument goes through, 
and this time we actually conclude that \begin{equation}
Z_{A(Q)'}[S^7,P_{[g_0]},H] = \exp\left(\frac{2\pi i }{12}\right).
\end{equation}

The classifying space $X$ of a $G$ gauge field and a 3-form field satisfying \eqref{dH=c2} 
 can be described as a total space of the fibration
\begin{equation}
	K(\bZ,3)
	\to
	X
	\to
	BG
\end{equation}
where the extension is controlled by \eqref{dH=c2}.
Therefore the relevant bordism group for us is $\Omega^\text{spin}_7(X)$.
See Appendix \ref{K(Z,3)-X-BG} for the details of the calculations;
we do find that $\Omega^\text{spin}_7(X)=\bZ_{12}$,
and our modified anomaly theory $A(Q)'$ detects its generator.
Again the cases $G=SU(3)$ and $G_2$ can be treated similarly,
and we will not detail it.

We now see that with 12 copies of the same fermion system, the global anomaly associated to $\pi_6(SU(2))$ vanishes,
when the manifolds involved are all equipped with a classical 3-form field $H$ satisfying \eqref{dH=c2}.
However, this \emph{does not} directly imply that one can perform 
the path integral over the $G$ gauge field and the 2-form field $B$ satisfying \eqref{dH=c2}.
Indeed, the $7d$ counterterm in \eqref{modification} with 12 copies corresponds to the $6d$ coupling \begin{equation}
- \dfrac{1}{2}\cdot B \left(c_2+\frac{p_1}2\right) 
\end{equation} 
but $\frac12(c_2+\frac{p_1}2)$ is not integrally quantized.
As we know from Sec.~\ref{sec:fermion}, we need to use chiral 2-form fields in $6d$
to cancel the fermion anomaly in this case,
and the subtlety of the chiral 2-forms is not encoded in the classical equation \eqref{dH=c2}.

%\newpage

Before closing, we note that the spacetime structure analogous to \eqref{dH=c2}, given by 
providing a 3-form field strength $H$ satisfying \begin{equation}
dH=\frac{p_1}{2}-c_2
\end{equation} on a spin manifold, is known as the \emph{(twisted) string structure} in mathematical literature.
This captures a part of the original $10d$ Green-Schwarz mechanism, where the fermion anomalies of the form \begin{equation}
\mathcal{I}_{12}=\left(\frac{p_1}2 -c_2\right) X_8
\end{equation}
is cancelled by the anomaly of the $B$-field.
Indeed, analogously to the analysis in this subsection,
equipping manifolds with (twisted) string structure automatically cancels the factorized $\mathcal{I}_{12}$.
It must be noted, however, that vanishing of the anomaly with the use of the (twisted) string structure
does not guarantee that the part $X_8$ is integrally quantized,
again as analogously to the $6d$ analysis.
This needs to be kept in mind whenever we try to use the string bordism to study the anomaly cancellation in string theory.

\section*{Acknowledgements}
The authors would like to thank the anonymous referee for the detailed comments,
which led to improve the representation of this paper.
Y.L.~is partially supported by the Programs for Leading Graduate Schools, MEXT, Japan, via the Leading Graduate Course for Frontiers of Mathematical Sciences and Physics
and also by JSPS Research Fellowship for Young Scientists.
Y.T.~is partially supported  by JSPS KAKENHI Grant-in-Aid (Wakate-A), No.17H04837 
and JSPS KAKENHI Grant-in-Aid (Kiban-S), No.16H06335,
and also by WPI Initiative, MEXT, Japan at IPMU, the University of Tokyo.

%\newpage
\bigskip

\appendix

%\newpage
\tikzset{every picture/.append style={scale=0.7},
every node/.append style={scale=0.9}}

\section{Fermion anomalies and group-theoretic constants}
\label{sec:group-theoretic}

The perturbative gauge anomaly of a fermion is given by
\begin{align}
	\cI_{\text{fermion}}
	&=
	\left[\widehat A(R) \ch(F)\right]_8\notag\\[10pt]
	&=
	(\mathrm{tr}_\text{rep.} 1) \cdot \frac{7\,p_1(R)^2-4\,p_2(R)}{5760}+
	\dfrac{p_1(R)}{24}
%	\cdot 
	\left[\dfrac{1}{2!}\cdot \mathrm{tr}_{\text{rep.}}\left(\frac{F}{2\pi}\right)^2\right]
	+ 
	\dfrac{1}{4!}\cdot \mathrm{tr}_{\text{rep.}}\left(\frac{F}{2\pi}\right)^4.
\end{align}
In general, the traces have following forms
\begin{equation}
	\renewcommand{\arraystretch}{1.3}
	\begin{array}{cclcc}
		\mathrm{tr}_{\text{rep.}}F^2
		& = &
		\alpha_{\text{rep.}}\cdot \mathrm{tr}_{\text{fund.}}F^2,\\
		\mathrm{tr}_{\text{rep.}}F^4
		& = &
		\beta_{\text{rep.}}\cdot \mathrm{tr}_{\text{fund.}}F^4
		& + & 
		\gamma_{\text{rep.}}\cdot \big(\mathrm{tr}_{\text{fund.}}F^2\big)^2,\\
	\end{array}
\end{equation}
and in particular $\beta_{\text{rep.}}=0$ for $SU(2)$, $SU(3)$, and $G_2$.
Explicitly, $\alpha$ and $\gamma$ for some common representations are given as follows:
\begin{equation}
	\renewcommand{\arraystretch}{2}
	\begin{array}{c|ccc|c}
		& SU(2) & SU(3) & G_2 & \\
		\hline
		\mathrm{tr}_{\text{adj.}}F^2 
		& 4 \cdot \mathrm{tr}_{\text{fund.}}F^2 
		& 6 \cdot \mathrm{tr}_{\text{fund.}}F^2 
		& 4 \cdot \mathrm{tr}_{\text{fund.}}F^2 & \alpha_{\text{adj.}}\\
		\mathrm{tr}_{\text{sym.}}F^2
		& -
		& 5 \cdot \mathrm{tr}_{\text{fund.}}F^2
		& - & \alpha_{\text{sym.}}\\
		\hline
		\mathrm{tr}_{\text{fund.}}F^4 
		& \dfrac{1}{2} \cdot \big(\mathrm{tr}_{\text{fund.}}F^2\big)^2
		& \dfrac{1}{2} \cdot \big(\mathrm{tr}_{\text{fund.}}F^2\big)^2
		& \dfrac{1}{4} \cdot \big(\mathrm{tr}_{\text{fund.}}F^2\big)^2 & \gamma_{\text{fund.}}\\
		\mathrm{tr}_{\text{adj.}}F^4
		& 8 \cdot \big(\mathrm{tr}_{\text{fund.}}F^2\big)^2
		& 9 \cdot \big(\mathrm{tr}_{\text{fund.}}F^2\big)^2
		& \dfrac{5}{2} \cdot \big(\mathrm{tr}_{\text{fund.}}F^2\big)^2 & \gamma_{\text{adj.}}\\
		\mathrm{tr}_{\text{sym.}}F^4 
		& -
		& \dfrac{17}{2} \cdot \big(\mathrm{tr}_{\text{fund.}}F^2\big)^2
		& - & \gamma_{\text{sym.}}\\
%		\hline
	\end{array}
\end{equation}
Furthermore, the instanton number $c_2(F)$ is normalized so that 
\begin{equation}
	\renewcommand{\arraystretch}{2}
	c_2(F) 
	= \dfrac{1}{4}\cdot \dfrac{1}{h^\vee} \cdot \mathrm{tr}_{\text{adj.}}\left(\frac{F}{2\pi}\right)^2
	=\left\{
		\begin{array}{cl}
			\dfrac{1}{4}\cdot \dfrac{1}{2}\cdot 4 \cdot \mathrm{tr}_{\text{fund.}}\left(\dfrac{F}{2\pi}\right)^2 = \dfrac{1}{2}\cdot \mathrm{tr}_{\text{fund.}}\left(\dfrac{F}{2\pi}\right)^2 & :SU(2),\\
			\dfrac{1}{4}\cdot \dfrac{1}{3}\cdot 6 \cdot \mathrm{tr}_{\text{fund.}}\left(\dfrac{F}{2\pi}\right)^2 = \dfrac{1}{2}\cdot \mathrm{tr}_{\text{fund.}}\left(\dfrac{F}{2\pi}\right)^2 & :SU(3),\\
			\dfrac{1}{4}\cdot \dfrac{1}{4}\cdot 4 \cdot \mathrm{tr}_{\text{fund.}}\left(\dfrac{F}{2\pi}\right)^2 = \dfrac{1}{4}\cdot \mathrm{tr}_{\text{fund.}}\left(\dfrac{F}{2\pi}\right)^2 & :G_2,\\
		\end{array}
	\right.
\end{equation}
where $h^\vee$ is the dual Coxeter number of each gauge group.
As a result, the non-gravitational part of the anomaly becomes
\begin{align}
	\label{anomaly-fermion}
	&\widetilde\cI_{\text{fermion}}
	=
	\dfrac{1}{24}\left(
		\alpha_{\text{rep.}}\cdot \mathrm{tr}_{\text{fund.}}\left(\frac{F}{2\pi}\right)^2\wedge \dfrac{p_1(R)}{2}
		+
		\gamma_{\text{rep.}}\cdot \mathrm{tr}_{\text{fund.}}\left(\frac{F}{2\pi}\right)^2\wedge \mathrm{tr}_{\text{fund.}}\left(\frac{F}{2\pi}\right)^2
	\right)\notag\\[6pt]
	&=
	\left\{
		\begin{array}{cccl}
			4\gamma_{\text{rep.}}\cdot \dfrac{1}{24}\cdot
				c_2(F)\left( c_2(F)
				+ \dfrac{p_1(R)}{2}
			\right)
			& + &
			\dfrac{2\alpha_{\text{rep.}}-4\gamma_{\text{rep.}}}{24}\cdot c_2(F) \wedge \dfrac{p_1(R)}{2}
			& : 
				SU(2),\\
			2\gamma_{\text{rep.}}\cdot \dfrac{1}{12}\cdot
				c_2(F)\left(c_2(F)+ \dfrac{p_1(R)}{2}
			\right)
			& + &
			\dfrac{2\alpha_{\text{rep.}}-4\gamma_{\text{rep.}}}{24}\cdot c_2(F) \wedge \dfrac{p_1(R)}{2}
			& : 	
				SU(3),
			\\ 
			4\gamma_{\text{rep.}}\cdot \dfrac{1}{6}\cdot
				c_2(F)\left(
				c_2(F)+ \dfrac{p_1(R)}{2}
			\right)
			& + &
			\dfrac{4\alpha_{\text{rep.}}-16\gamma_{\text{rep.}}}{24}\cdot c_2(F) \wedge \dfrac{p_1(R)}{2}
			& : \begin{array}{l}
				G_2.
			\end{array}
		\end{array}
	\right.
\end{align}
From our general argument in Sec.~\ref{sec:SU2}, \ref{sec:SU3} and \ref{sec:G2},
 the numerical coefficients before $\cdot$ of the two terms in each row,
such as $4\gamma_\text{rep}$ and $\frac1{24}(2\alpha_\text{rep.}-4\gamma_\text{rep.})$,
are guaranteed to be integers.

\newpage

\section{Bordisms via Atiyah-Hirzebruch spectral sequence}
\label{sec:AHSS}
In this appendix, we compute $\Omega^\text{spin}_d(X)$ for $X=K(\bZ,4)$, $BSU(2)$, $BSU(3)$, $BG_2$, 
and $K(\bZ,3)$-fibered $BSU(2)$
by using the Atiyah-Hirzebruch spectral sequence (AHSS),
associated to the trivial fibration
\begin{equation}
	pt
	\longrightarrow
	X
	\overset{p}{\longrightarrow}
	X.
\end{equation}
Here we will not give an introduction to AHSS; consult e.g.~\cite{Garcia-Etxebarria:2018ajm} for a detailed introduction to AHSS for $\Omega^\text{spin}_*$.\footnote{
	We note that the AHSS appeared in the physics literature first in the context of K-theory classification of D-branes in \cite{Diaconescu:2000wy,Bergman:2001rp,Maldacena:2001xj};
	see Appendix C, Appendix A, Sec.~3 of respective references.
}

\subsection{$K(\bZ,4)$}
The $E^2$-page of the AHSS is as follows:
\begin{equation}
	\label{K(Z,4)-AHSS}
	\begin{array}{ccc}
		E^2_{p,q}=H_p\big(K(\bZ,4);\Omega^{\text{spin}}_q\big) && \widetilde \Omega_{p+q}^{\text{spin}}(K(\bZ,4))\vspace{2mm}\\
		\begin{array}{c|c:ccccccccc}
			8  & \bZ^{\oplus 2} &&&& \ast && \ast && \ast &\\
			7  & \cellcolor{lightyellow}\\
			6  & & \cellcolor{lightyellow}\\
			5  & && \cellcolor{lightyellow}\hphantom{\bZ_2} & \hphantom{\bZ_2} & \\
			4  & \bZ &&& \cellcolor{lightyellow} & \bZ && \bZ_2 && \ast &\\
			3  & &&&& \cellcolor{lightyellow}\\
			2  & \bZ_2 &&&& \Blue{\fbox{\black{$\bZ_2$}}} & \cellcolor{lightyellow} & \bZ_2 & \bZ_2& \bZ_2 & \\
			1  & \bZ_2 & \hphantom{\bZ_2} & \hphantom{\bZ_2} & \hphantom{\bZ_2} & \red{\fbox{\black{$\bZ_2$}}}& \hphantom{\bZ_2} & \Blue{\fbox{\black{$\bZ_2$}}}\cellcolor{lightyellow} & \bZ_2 & \bZ_2& \hphantom{\bZ_2} \\
			0 & \bZ &&&& \bZ && \red{\fbox{\black{$\bZ_2$}}} & \cellcolor{lightyellow} & \bZ\oplus \bZ_3 & \\
			\hline
			& 0 & 1 & 2 & 3 & 4 & 5 & 6 & 7 & 8 & 9\\
		\end{array}
		& \Longrightarrow & 
		\begin{array}{c|c}
			8  & \ast\\
			7  & \cellcolor{lightyellow}\\
			6  & \\
			5  & \\
			4  & \bZ\\
			3  & \\
			2  & \\
			1  & \\
			0 & \\
			\hline\\
		\end{array}
	\end{array}
\end{equation}
where the horizontal axis and the vertical axis are for $p$ and $q$, respectively.
The integral homology of $K(\bZ,4)$ is due to \cite[Appendix B.2]{BMT13}. 
For $\bZ_2$ (co)homology, it is known that (see e.g. \cite{MimuraToda})
\begin{equation}
	H^\ast(K(\bZ,4);\bZ_2) = \bZ_2[u_4, u_6, u_7, u_{10}, u_{11}, u_{13}, \cdots]
\end{equation}
where
\begin{equation}
	\begin{array}{lcc}
		Sq^2 u_4 & = & u_6,\\
		Sq^1 u_6 & = & u_7.\\
	\end{array}
\end{equation}
The differentials 
\red{\fbox{\black{$d_2:E^2_{6,0}\to E^2_{4,1}$}}} and 
\Blue{\fbox{\black{$d_2:E^2_{6,1}\to E^2_{4,2}$}}} are known to be 
(mod 2 reduction composed with) the dual of $Sq^2$\cite{Teichner93},
and therefore none of the $\bZ_2$'s involved survive to the $E^\infty$-page.\footnote{
	This indeed agrees with the result of \cite{StongAppendix} up to odd torsion.
}
As a result, one has $\widetilde\Omega^{\text{spin}}_7(K(\bZ,4))=0$.
The extension problem for $\widetilde\Omega^{\text{spin}}_8(K(\bZ,4))$ was solved in Sec.~\ref{sec:SU3},
see in particular the discussions around \eqref{extension-problem}.
Summarizing, we have the (non-canonical) isomorphism of groups
\begin{equation}
	\Inv_{\text{spin}}^7(K(\bZ,4)) 
	= 
	\underbrace{\Free\widetilde\Omega^{\text{spin}}_8(K(\bZ,4))}_{=\bZ\oplus \bZ}
	\oplus 
	\underbrace{\Tors\widetilde\Omega^{\text{spin}}_7(K(\bZ,4))}_{=0}.
\end{equation}

\subsection{$BSU(2)$}
The $E^2$-page of the AHSS is as follows:
\begin{equation}
	\begin{array}{ccc}
		E^2_{p,q}=H_p\big(BSU(2);\Omega^{\text{spin}}_q\big) && \widetilde \Omega_{p+q}^{\text{spin}}(BSU(2))\vspace{2mm}\\
		\begin{array}{c|c:cccccccccccc}
			8  & \bZ^{\oplus 2} &&&& \ast &&&& \ast\\
			7  & \cellcolor{lightyellow}\\
			6  & & \cellcolor{lightyellow}\\
			5  &&& \cellcolor{lightyellow}\\
			4  & \bZ &&& \cellcolor{lightyellow} & \bZ &&&& \ast\\
			3  &&&&& \cellcolor{lightyellow} &&&& \\
			2  & \bZ_2 &&&& \bZ_2 & \cellcolor{lightyellow} &&& \ast &\\
			1  & \bZ_2 & \hphantom{\bZ_2} & \hphantom{\bZ_2} & \hphantom{\bZ_2} & \bZ_2 & \hphantom{\bZ_2} & \hphantom{\bZ_2}\cellcolor{lightyellow} & \hphantom{\bZ_2} & \ast & \hphantom{\bZ_2}\\
			0 & \bZ &&&& \bZ &&& \cellcolor{lightyellow} & \bZ & \\
			\hline
			& 0 & 1 & 2 & 3 & 4 & 5 & 6 & 7 & 8 & 9 \\
		\end{array}
		& \Longrightarrow & 
		\begin{array}{c|c}
			8  & \bZ\oplus\bZ\\
			7  & \cellcolor{lightyellow}\\
			6  & \bZ_2\\
			5  & \bZ_2\\
			4  & \bZ\\
			3  & \\
			2  & \\
			1  & \\
			0 & \\
			\hline\\
		\end{array}
	\end{array}
\end{equation}
where the relevant (co)homologies are
\begin{equation}
	\renewcommand{\arraystretch}{1.2}
	\begin{array}{ccc}
		H^\ast\big(BSU(2);\bZ\big) & = & \bZ[c_2],\\
		H^\ast\big(BSU(2);\bZ_2\big) & = & \bZ_2[c_2].
	\end{array}
\end{equation}
For the range $p+q\leq 7$, there are no differentials and all the elements survive to $E^\infty$-page.
As a result, one has
\begin{equation}
	\Inv_{\text{spin}}^7(BSU(2)) 
	= 
	\underbrace{\Free\widetilde\Omega^{\text{spin}}_8(BSU(2))}_{=\bZ\oplus \bZ}
	\oplus 
	\underbrace{\Tors\widetilde\Omega^{\text{spin}}_7(BSU(2))}_{=0}.
\end{equation}
Since $BSU(2)=\HP^\infty$, 
each generator of $H_{4i}(BSU(2);\bZ)=\bZ$ is the embedded $\HP^{4i}$.
As $SU(2)=Sp(1)$ bundles, they are the canonical bundles $Q$ over $\HP^{4i}$.
Therefore, the generator of $E^{2}_{4,4}=\bZ$ is $(\HP^1,Q)\times K3$
and the generator of $E^2_{8,0}=\bZ$ is $(\HP^2,Q)$.
They are the generators discussed in Sec.~\ref{sec:SU2}.

\subsection{$BSU(3)$}
The $E^2$-page of the AHSS is as follows:
\begin{equation}
	\begin{array}{ccc}
		E^2_{p,q}=H_p\big(BSU(3);\Omega^{\text{spin}}_q\big) && \widetilde \Omega_{p+q}^{\text{spin}}(BSU(3))\vspace{2mm}\\
		\begin{array}{c|c:cccccccccccc}
			8  & \bZ^{\oplus 2} &&&& \ast && \ast && \ast\\
			7  & \cellcolor{lightyellow}\\
			6  & & \cellcolor{lightyellow}\\
			5  &&& \cellcolor{lightyellow}\\
			4  & \bZ &&& \cellcolor{lightyellow} & \bZ && \ast && \ast\\
			3  &&&&& \cellcolor{lightyellow} &&&& \\
			2  & \bZ_2 &&&& \Blue{\fbox{\black{$\bZ_2$}}} & \cellcolor{lightyellow} & \bZ_2 && \ast &\\
			1  & \bZ_2 & \hphantom{\bZ_2} & \hphantom{\bZ_2} & \hphantom{\bZ_2} & \red{\fbox{\black{$\bZ_2$}}} & \hphantom{\bZ_2} & \Blue{\fbox{\black{$\bZ_2$}}}\cellcolor{lightyellow} & \hphantom{\bZ_2} & \ast & \hphantom{\bZ_2}\\
			0 & \bZ &&&& \bZ && \red{\fbox{\black{$\bZ$}}} & \cellcolor{lightyellow} & \bZ & \\
			\hline
			& 0 & 1 & 2 & 3 & 4 & 5 & 6 & 7 & 8 & 9 \\
		\end{array}
		& \Longrightarrow & 
		\begin{array}{c|c}
			8  & \ast\\
			7  & \cellcolor{lightyellow}\\
			6  & \bZ\\
			5  & \\
			4  & \bZ\\
			3  & \\
			2  & \\
			1  & \\
			0 & \\
			\hline\\
		\end{array}
	\end{array}
\end{equation}
where the relevant (co)homologies are
\begin{equation}
	\renewcommand{\arraystretch}{1.2}
	\begin{array}{ccc}
		H^\ast\big(BSU(3);\bZ\big) & = & \bZ[c_2, c_3],\\
		H^\ast\big(BSU(3);\bZ_2\big) & = & \bZ_2[c_2, c_3].
	\end{array}
\end{equation}
Here, $c_2$ and $c_3$ in $H^\ast\big(BSU(3);\bZ_2\big)$ are related as
\begin{equation*}
	Sq^2 c_2 = c_3,
\end{equation*}
and therefore the differentials 
\red{\fbox{\black{$d_2:E^2_{6,0}\to E^2_{4,1}$}}} and 
\Blue{\fbox{\black{$d_2:E^2_{6,1}\to E^2_{4,2}$}}} are again non-trivial
%(mod 2 reduction composed with) the dual of $Sq^2$\cite{Teichner93},
as in the $X=K(\bZ,4)$ case.
As a result, one has
\begin{equation}
	\Inv_{\text{spin}}^7(BSU(3)) 
	= 
	\underbrace{\Free\widetilde\Omega^{\text{spin}}_8(BSU(3))}_{=\bZ\oplus \bZ}
	\oplus 
	\underbrace{\Tors\widetilde\Omega^{\text{spin}}_7(BSU(3))}_{=0}.
\end{equation}
As discussed in Sec.~\ref{sec:SU3},
$E^2_{6,2}=\bZ_2$  extends $E^2_{4,4}=\bZ$ to form a $\bZ$.
The generator is a dual basis to $\tilde I_{\mathbf{3}}= \frac16 \cdot \frac12 c_2(c_2+\frac{p_1}2)$.
Another factor of $\bZ$ simply comes from $H_8(BSU(3);\bZ)=\bZ$.

\subsection{$BSU(n\geq 4)$}
In passing, we briefly comment on the case of $BSU(n\geq 4)$.
The (relevant part of) $E^2$-page of the AHSS is almost the same as $BSU(3)$.
The only difference is the additional $\bZ$ in $E^2_{8,\ast}$ elements due to $c_4\in H^8(BSU(n);\bZ)$,
which also newly shows up in the anomaly polynomial $\tilde\cI_{\mathbf n}$.
As a result, one has
\begin{equation}
	\Inv_{\text{spin}}^7(BSU(n)) 
	= 
	\underbrace{\Free\widetilde\Omega^{\text{spin}}_8(BSU(n))}_{=\bZ\oplus \bZ\oplus\bZ}
	\oplus 
	\underbrace{\Tors\widetilde\Omega^{\text{spin}}_7(BSU(n))}_{=0}.
\end{equation}

Comparing with the $BSU(3)$ case, we find that one factor of $\bZ$ is given by extending $E^{2}_{4,4}=\bZ$ by $E^2_{6,2}=\bZ_2$ as before, whose generator is obtained by sending the corresponding one for $BSU(3)$ by the embedding $SU(3)\to SU(n)$. This is a dual basis to 
$
\tilde \cI_{\mathbf{n}}=-\frac{c_4}6 + \frac{1}{6}\cdot\frac{1}{2}c_2(c_2+\frac{p_1}2).
$
The other two factors of $\bZ$ simply come from $H_8(BSU(n);\bZ)=\bZ\oplus\bZ$
and are dual  to $(c_2)^2$ and $c_4$.
Then we see that a basis of $\Hom_\bZ(\widetilde\Omega^\text{spin}_8(BSU(n)),\bZ)$ can be chosen to be
$\tilde\cI_{\mathbf{n}}$, $(c_2)^2$ and $\frac12c_2(c_2+\frac{p_1}2)$, 
the last two of which can be obtained by pulling back from $K(\bZ,4)$.
Applying the logic of Sec.~\ref{sec:fermion}, we find that
the anomaly of 2-form fields can cancel the fermion anomaly if and only if the coefficient of $c_4$ in the anomaly polynomial vanishes.

\subsection{$BG_2$}
The $E^2$-page of the AHSS is as follows:
\begin{equation}
	\label{BG2-AHSS}
	\begin{array}{ccc}
		E^2_{p,q}=H_p\big(BG_2;\Omega^{\text{spin}}_q\big) && \widetilde \Omega_{p+q}^{\text{spin}}(BG_2)\vspace{2mm}\\
		\begin{array}{c|c:cccccccccccc}
			8  & \bZ^{\oplus 2} &&&& \ast && \ast && \ast\\
			7  & \cellcolor{lightyellow}\\
			6  & & \cellcolor{lightyellow}\\
			5  &&& \cellcolor{lightyellow}\\
			4  & \bZ &&& \cellcolor{lightyellow} & \bZ && \ast && \ast\\
			3  &&&&& \cellcolor{lightyellow} &&&& \\
			2  & \bZ_2 &&&& \Blue{\fbox{\black{$\bZ_2$}}} & \cellcolor{lightyellow} & \bZ_2 & \bZ_2 & \ast &\\
			1  & \bZ_2 & \hphantom{\bZ_2} & \hphantom{\bZ_2} & \hphantom{\bZ_2} & \red{\fbox{\black{$\bZ_2$}}} & \hphantom{\bZ_2} & \Blue{\fbox{\black{$\bZ_2$}}}\cellcolor{lightyellow} & \bZ_2 & \ast & \hphantom{\bZ_2}\\
			0 & \bZ &&&& \bZ && \red{\fbox{\black{$\bZ_2$}}} & \cellcolor{lightyellow} & \bZ & \\
			\hline
			& 0 & 1 & 2 & 3 & 4 & 5 & 6 & 7 & 8 & 9 \\
		\end{array}
		& \Longrightarrow & 
		\begin{array}{c|c}
			8  & \ast\\
			7  & \cellcolor{lightyellow}\\
			6  & \\
			5  & \\
			4  & \bZ\\
			3  & \\
			2  & \\
			1  & \\
			0 & \\
			\hline\\
		\end{array}
	\end{array}
\end{equation}
where the relevant (co)homologies are
\begin{equation}
	\renewcommand{\arraystretch}{1.2}
	\begin{array}{lcl}
		H^\ast\big(BG_2;\bZ_{p\geq 3}\big) & = & \bZ_p[x_4, x_{12}],\\
		H^\ast\big(BG_2;\bZ_2\big) & = & \bZ_2[x_4, x_6, x_7].
	\end{array}
\end{equation}
For the latter, generators are related as
\begin{equation}
	\renewcommand{\arraystretch}{1.1}
	\begin{array}{lcc}
		Sq^2 x_4 & = & x_6,\\
		Sq^1 x_6 & = & x_7,\\
	\end{array}
\end{equation}
and therefore the differentials 
\red{\fbox{\black{$d_2:E^2_{6,0}\to E^2_{4,1}$}}} and 
\Blue{\fbox{\black{$d_2:E^2_{6,1}\to E^2_{4,2}$}}} are again non-trivial
as in the $X=K(\bZ,4)$ case.
%Combining with the knowledge of $E^\infty$-page which will be obtained later in Appendix \ref{BG2-Adams},
Since the (co)homology of $BG_2$ is $p$-torsion free for $p\geq 3$ (see e.g. \cite{MimuraToda}),
one can deduce the integral (co)homology as in the $E^2$-page.
As a result, one has
\begin{equation}
	\Inv_{\text{spin}}^7(BG_2) 
	= 
	\underbrace{\Free\widetilde\Omega^{\text{spin}}_8(BG_2)}_{=\bZ\oplus \bZ}
	\oplus 
	\underbrace{\Tors\widetilde\Omega^{\text{spin}}_7(BG_2)}_{=0}.
\end{equation}

\subsection{$K(\bZ,3)\to X\to BSU(2)$}
\label{K(Z,3)-X-BG}
Here we compute $\Omega_d^\text{spin}(X)$ 
for a fibration $K(\bZ,3)\to X\to BSU(2)$ designed to kill the generator $c_2\in H^4(BSU(2);\bZ)=\bZ$
via $dH=c_2$, where $H\in H^3(K(\bZ,3);\bZ)=\bZ$ is also the generator.
To compute $\Omega_d^{\text{spin}}(X)$, 
let us first calculate the cohomology group of $X$ via the Leray-Serre spectral sequence (LSSS),
and then throw it into the AHSS.
%\subsubsection{$G=SU(2)$}
The $E_2$-page of the LSSS is given on the left hand side of the following equation:
\begin{equation}
	\label{X-SU(2)-LSSS}
	\begin{array}{ccc}
		E_2^{p,q}=H^p\big(BSU(2);H^q(K(\bZ,3);\bZ)\big) && H^{p+q}(X;\bZ)\vspace{2mm}\\
		\begin{array}{c|cccccccccc}
			9  & \red{\dbox{\black{$\bZ_2$}}} &&&& \ast &&&& \ast & \\
			8  & \bZ_3 &&&& \ast &&&& \ast & \\
			7  & \\
			6  & \bZ_2&  &&& \red{\dbox{\black{$\bZ_2$}}} &&&& \ast\\
			5  & && \hphantom{\bZ_2} & \hphantom{\bZ_2} & \\
			4  & &&&  &&&&&&\\
			3  & \red{\fbox{\black{$\bZ$}}} &&&& \Blue{\fbox{\black{$\bZ$}}} &&&& \ast\\
			2  & &&&&&  &&&& \\
			1  & & \hphantom{\bZ_2} & \hphantom{\bZ_2} & \hphantom{\bZ_2} & \hphantom{\bZ_2} & \hphantom{\bZ_2} & \hphantom{\bZ_2} & \hphantom{\bZ_2} & \hphantom{\bZ_2} & \hphantom{\bZ_2} \\
			0 & \bZ &&&& \red{\fbox{\black{$\bZ$}}} &&&  & \Blue{\fbox{\black{$\bZ$}}} & \\
			\hline
			& 0 & 1 & 2 & 3 & 4 & 5 & 6 & 7 & 8 & 9\\
		\end{array}
		& \Longrightarrow & 
		\begin{array}{c|c}
			9 & \\
			8  & \bZ_3\\
			7  & \\
			6  & \bZ_2\\
			5  & \\
			4  & \\
			3  & \\
			2  & \\
			1  & \\
			0 & \bZ\\
			\hline\\
		\end{array}
	\end{array}
\end{equation}
where the integral homology of $K(\bZ,3)$ is again due to \cite[Appendix B.2]{BMT13}.
The differentials \red{\fbox{\black{$d_4:E_2^{0,3}\to E_2^{4,0}$}}} and \red{\dbox{\black{$d_4:E_2^{0,9}\to E_2^{4,6}$}}}
are non-trivial due to $dH=c_2$,
and \Blue{\fbox{\black{$d_4:E_2^{4,3}\to E_2^{8,0}$}}} is also non-trivial due to
\begin{equation}
	d(H\wedge c_2(F)) = (c_2(F))^2.
\end{equation}
This results in the cohomology group of $X$ as given on the right hand side of \eqref{X-SU(2)-LSSS}.

The knowledge of $H^*(X,\bZ)$ allows us to compute the AHSS for $pt\to X\to X$,
whose $E^2$-page is as follows:
\begin{equation}
	\label{X-SU(2)-AHSS}
	\begin{array}{c}
		E^2_{p,q}=H_p\big(X;\Omega^{\text{spin}}_q\big) \vspace{2mm}\\%&& \widetilde \Omega_{p+q}^{\text{spin}}(X)\vspace{2mm}\\
		\begin{array}{c|c:ccccccccc}
			8  & \bZ^{\oplus 2} &&&&& \ast && \ast & \\
			7  & \cellcolor{lightyellow}\\
			6  & & \cellcolor{lightyellow}\\
			5  & && \cellcolor{lightyellow}\\
			4  & \bZ &&& \cellcolor{lightyellow} && \bZ_2 && \ast & \\
			3  & &&&& \cellcolor{lightyellow} &&&& \hphantom{\bZ_2}\\
			2  & \bZ_2 &&&&& \bZ_2\cellcolor{lightyellow} & \bZ_2 && \\
			1  & \bZ_2 & \hphantom{\bZ_2} & \hphantom{\bZ_2} & \hphantom{\bZ_2} & \hphantom{\bZ_2} & \bZ_2 & \bZ_2\cellcolor{lightyellow} & \hphantom{\bZ_2} & \\
			0 & \bZ &&&&& \bZ_2 && \bZ_3\cellcolor{lightyellow} & \\
			\hline
			& 0 & 1 & 2 & 3 & 4 & 5 & 6 & 7 & 8\\
		\end{array}
%		& \Longrightarrow & 
%		\begin{array}{c|c}
%			8  & \ast\\
%			7  & \cellcolor{lightyellow}\\
%			6  & \\
%			5  & \\
%			4  & \bZ\\
%			3  & \\
%			2  & \\
%			1  & \\
%			0 & \\
%			\hline\\
%		\end{array}
	\end{array}
\end{equation}
where the relevant homology groups can be obtained from the universal coefficient theorems.
As a result, one has
\begin{equation}
	\Inv_{\text{spin}}^7(X) 
	= 
	\underbrace{\Free\widetilde\Omega^{\text{spin}}_8(X)}_{=0}
	\oplus 
	\Tors\widetilde\Omega^{\text{spin}}_7(X)
\end{equation}
and $\Tors\widetilde\Omega^{\text{spin}}_7(X)$ is a finite group with at most $2\cdot 2\cdot 3=12$ elements.
As we saw in Sec.~\ref{sec:take2},
there is a bordism invariant with the value $\exp(2\pi i/12)$, meaning that 
the AHSS is indeed consistent with the previous analyses on $6d$ global anomaly claiming
\begin{equation}
	\widetilde\Omega^{\text{spin}}_7(X)=\bZ_{12}.
\end{equation}

\newpage

\section{Bordisms  via Adams spectral sequence}
\label{sec:Adams}
In this appendix, we compute $\Omega^\text{spin}_{d}(BG)$ for $G=SO(n)$, $Spin(n)$ and $G_2$ 
by using the Adams spectral sequence
\begin{equation}
	E_2^{s,t} 
	= \mathrm{Ext}_{\cA(1)}^{s,t}(\widetilde{H}^*(X;\mathbb{Z}_2),\mathbb{Z}_2)
	\quad
	\Rightarrow
	\quad
	\widetilde{\mathrm{ko}}_{t-s}(X)_2^{\wedge}
	\label{eq:ASS}
\end{equation}
where 
$\cA(1)$ is the algebra generated by the Steendrod operations $Sq^1$ and $Sq^2$,
and
$\mathrm{ko}$ is the connective KO theory.
The RHS is known to agree with the reduced spin-bordism $\widetilde\Omega_{t-s}^{\text{spin}}(X)$ for $t-s\leq 7$\cite{ABP1967},
allowing us to compute the spin bordism groups of our interest.
Here we will not give an introduction to Adams SS; consult \cite{BCGuide} for the details
(to which we refer for the Adams charts of various modules below),
or see \cite[Appendix C]{Lee:2020ojw} for a brief description.

\subsection{$BSO(n)$}
\paragraph{$n\geq 5\ :$} 
The module $M^{BSO}=\widetilde{H}^*(BSO(n);\mathbb{Z}_2)$ up to degree $7$ is as follows:
\begin{equation}
	\begin{tikzpicture}[thick,baseline=(m-4-2)]
		\matrix (m) [matrix of math nodes, row sep= 1em,
		column sep=6em, ]{
			\phantom{\bullet} & \bullet & \bullet & \bullet & \bullet\\
		    \bullet& \bullet & \phantom{\bullet} & \bullet & \bullet\\
			\bullet & \bullet \\
			\bullet & \bullet \\
			\bullet \\
			\bullet \\
		};
		\draw (m-1-3.center) to [out=260, in =80] (m-3-2.center);
		\draw (m-1-2.center) to (m-2-2.center);
		\draw (m-1-4.center) to (m-2-4.center);
		\draw (m-1-5.center) to (m-2-5.center);
		\draw (m-2-2.center) to [out=220, in = 140] (m-4-2.center);
		\draw (m-3-2.center) -- (m-4-2.center);
		\draw (m-3-1.center) to [out=320,in=50] (m-5-1.center);
		\draw (m-2-1.center) to [out=220,in=140] (m-4-1.center);
		\draw (m-4-1.center) to [out=220,in=140] (m-6-1.center);
		\draw (m-2-1.center) to (m-3-1.center);
		\draw (m-6-1.center) -- (m-5-1.center);
		\node[anchor = west] at (m-6-1) {$w_2$};
		\node[anchor = west] at (m-5-1) {$w_3$};
		\node[anchor = west] at (m-4-2) {$w_4$};
		\node[anchor = west] at (m-3-1) {$w_2 w_3+w_5$};
		\node[anchor = west] at (m-3-2) {$w_5$};
		\node[anchor = east] at (m-2-2) {$w_2w_4+w_6$};
		\node[anchor = west] at (m-1-3) {$w_2w_5$};
		\node[anchor = east] at (m-1-2) {$w_2w_5+w_3w_4+w_7$};
		\node[anchor = west] at (m-1-4) {$w_7$};
		\node[anchor = west] at (m-2-4) {$w_6$};
		\node[anchor = west] at (m-1-5) {$w_3(w_2)^2$};
		\node[anchor = west] at (m-2-5) {$(w_2)^3$};
		\node[anchor = east] at (m-4-1) {$(w_2)^2$};
		\node[anchor = east] at (m-2-1) {$(w_3)^2$};
	\end{tikzpicture}
\end{equation}
where, as is customary, we used a dot for a basis in a $\bZ_2$-vector space,
and a vertical straight line and a curved line shows the action of $Sq^1$ and $Sq^2$.
We therefore have
\begin{equation}
	M^{BSO}_{\le 7} \cong \big(J[2]\oplus \cA(1)[4]\oplus \cA(1)[6]\oplus \cA(1)[6]\big)_{\le 7}
\end{equation}
in terms of \cite{BCGuide}, and the Adams chart is correspondingly given as
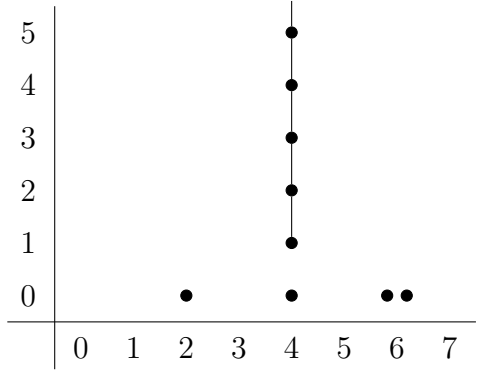
\begin{center}
\DeclareSseqGroup\tower {} {
	\class(0,0)\foreach \i in {1,...,8} {
		\class(0,\i)
		\structline(0,\i-1,-1)(0,\i,-1)
	}
}
\begin{sseqdata}[name=MBSO,Adams grading,classes = fill,xrange = {0}{7},yrange = {0}{5}
%,x label = {$t-s$}, y label = {$s$}
]
	\class(2,0)
%	\class(3,1)
%	\structline(2,0)(3,1)
	\tower(4,1)
	\class(4,0)
	\class(6,0)
	\class(6,0)
\end{sseqdata}
\printpage[name = MBSO,page = 2]
\end{center}
where the horizontal axis and the vertical axis are for $t-s$ and $s$, 
a dot corresponds to a generator,
and the vertical line shows the action of a multiplication by a special element known as $h_0$;
a tower of $h_0$ generates the ring $\bZ_{2}$ of $2$-adic integers.
As this $E_2$ page is too sparse for any differentials, one obtains
\begin{equation}
	\renewcommand{\arraystretch}{1.3}
	\begin{array}{c|cccccccc}
		d & 0 & 1 & 2 & 3 & 4 & 5 & 6 & 7\\
		\hline
		\widetilde\Omega_{d}^{\text{spin}}(BSO(n\geq 5))
		& 0 & 0 & \bZ_2 & 0 & \bZ\oplus \bZ_2 & 0 & \bZ_2^{\oplus 2} & 0\\
		\hline
	\end{array}
\end{equation}

\paragraph{$n=4\ :$} 
Removing $w_{i\geq 5}$ from the module $M^{BSO}$, it becomes $(J[2]\oplus Q[4]\oplus \cA(1)[6])_{\le 7}$ as $\cA(1)$-module.
Correspondingly, the Adams chart becomes
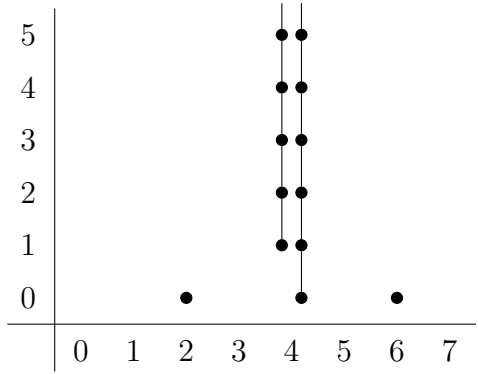
\begin{center}
\begin{sseqdata}[name=MBSO2,Adams grading,classes = fill,xrange = {0}{7},yrange = {0}{5}%,x label = {$t-s$}, y label = {$s$}
]
	\class(2,0)
%	\class(3,1)
%	\structline(2,0)(3,1)
	\tower(4,1)
	\class[white](4,0)
	\tower(4,0)
	\class(6,0)
\end{sseqdata}
\printpage[name = MBSO2,page = 2]
\end{center}
Again, this $E_2$ page is too sparse for any differentials, and one obtains
\begin{equation}
	\renewcommand{\arraystretch}{1.3}
	\begin{array}{c|cccccccc}
		d & 0 & 1 & 2 & 3 & 4 & 5 & 6 & 7\\
		\hline
		\widetilde\Omega_{d}^{\text{spin}}(BSO(4))
		& 0 & 0 & \bZ_2 & 0 & \bZ^{\oplus 2} & 0 & \bZ_2 & 0\\
		\hline
	\end{array}
\end{equation}

\paragraph{$n=3\ :$} 
Further removing $w_{4}$ from the module $M^{BSO}$, it becomes $(J[2]\oplus \cA(1)[6])_{\le 7}$ as $\cA(1)$-module.
Correspondingly, the Adams chart becomes
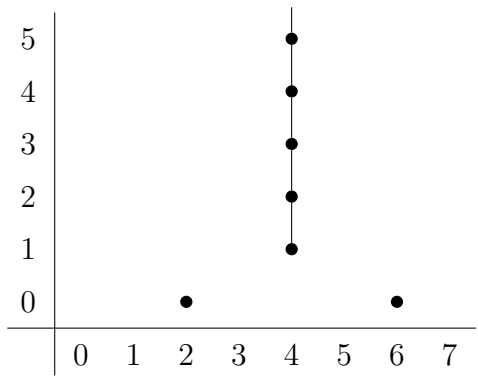
\begin{center}
\begin{sseqdata}[name=MBSO3,Adams grading,classes = fill,xrange = {0}{7},yrange = {0}{5}%,x label = {$t-s$}, y label = {$s$},title= $E_2$
]
	\class(2,0)
%	\class(3,1)
%	\structline(2,0)(3,1)
	\tower(4,1)
	\class(6,0)
\end{sseqdata}
\printpage[name = MBSO3,page = 2]
\end{center}
Again, this $E_2$ page is too sparse for any differentials, and one obtains
\begin{equation}
	\renewcommand{\arraystretch}{1.3}
	\begin{array}{c|cccccccc}
		d & 0 & 1 & 2 & 3 & 4 & 5 & 6 & 7\\
		\hline
		\widetilde\Omega_{d}^{\text{spin}}(BSO(3))
		& 0 & 0 & \bZ_2 & 0 & \bZ & 0 & \bZ_2 & 0\\
		\hline
	\end{array}
\end{equation}

\subsection{$BSpin(n)$}
For $n\leq 6$, one can use the exceptional isomorphisms to deduce $\widetilde\Omega_{7}^{\text{spin}}(BSpin(n))=0$.
Therefore, let us focus on the remaining $n\geq 7$ cases.
Since $M^{BSpin}=H^\ast(BSpin(n);\bZ_2)$ up to degree 7 
can be obtained by removing $w_2$, $w_3$, and $w_5$ from $M^{BSO}_{\leq 7}$ \cite[Theorem~6.5]{QuillenBSpin}\footnote{
	This happens to be completely the same as $K(\bZ,4)$, 
	and the result is indeed consistent with the AHSS computation \eqref{K(Z,4)-AHSS}. 
	See also \cite{JohnFrancis2005}.
}, it is
\begin{equation}
	\begin{tikzpicture}[thick,baseline=(m-2-1.south)]
		\matrix (m) [matrix of math nodes, row sep= 1em,
		column sep=6em, ]{
			\bullet\\
		    \bullet\\
			\phantom{\bullet}\\
			\bullet \\
		};
		\draw (m-1-1.center) to (m-2-1.center);
		\draw (m-2-1.center) to [out=220, in = 140] (m-4-1.center);
		\node[anchor = west] at (m-4-1) {$w_4$};
		\node[anchor = west] at (m-2-1) {$w_6$};
		\node[anchor = west] at (m-1-1) {$w_7$};
	\end{tikzpicture}
\end{equation}
which means
\begin{equation}
	M^{BSpin}_{\le 7} \cong Q[4].
\end{equation}
Therefore, the Adams chart is
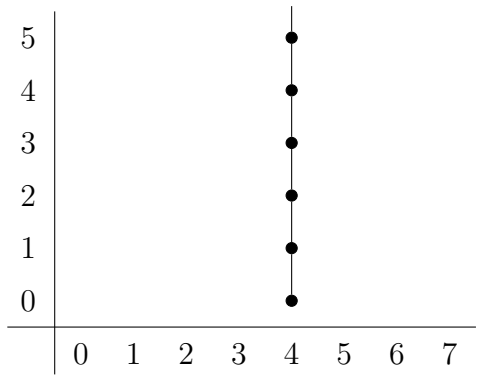
\begin{center}
\begin{sseqdata}[name=MBSpin,Adams grading,classes = fill,xrange = {0}{7},yrange = {0}{5}%,x label = {$t-s$}, y label = {$s$}
]
	\tower(4,0)
\end{sseqdata}
\printpage[name = MBSpin,page = 2]
\end{center}
As there is no room for any differentials, one obtains
\begin{equation}
	\label{omegaBSpin}
	\renewcommand{\arraystretch}{1.3}
	\begin{array}{c|cccccccc}
		d & 0 & 1 & 2 & 3 & 4 & 5 & 6 & 7\\
		\hline
		\widetilde\Omega_{d}^{\text{spin}}(BSpin(n\geq 7))
		& 0 & 0 & 0 & 0 & \bZ & 0 & 0 & 0\\
		\hline
	\end{array}
\end{equation}

\subsection{$BG_2$}
\label{BG2-Adams}
Since $H^*(BG_2;\bZ_2) =\bZ_2[x_4, x_6, x_7]$ 
and each elements are related as \cite{MimuraToda}
\begin{equation}
	\begin{array}{lcc}
		Sq^2 x_4 & = & x_6,\\
		Sq^1 x_6 & = & x_7,\\
	\end{array}
\end{equation}
the result is completely the same as the $BSpin$ case, and in particular one has $\widetilde\Omega_{7}^{\text{spin}}(BG_2) = 0$.
This is also consistent with the AHSS computation \eqref{BG2-AHSS}.

\newpage

\section{Recollections of Eguchi-sensei}
\label{sec:memory}

In this appendix, one of the authors (Yuji Tachikawa) would like to share memories of his late advisor Professor Tohru Eguchi, who passed away unexpectedly in early 2019. 
In the rest of the appendix, I would like to use the first person, and I would also like to refer to my late advisor as Eguchi-sensei (江口先生) as I often called him.

My first encounter with him should have been when I was reading the August 1994 issue of a popular Japanese mathematics magazine, ``Suugaku seminar'', 
in which there was a write-up of his interview of Professor Edward Witten.
It was interesting for me to re-read that article to prepare this appendix. 
The interview was dated May 13, 1994, and there was a discussion of the Seiberg-Witten theory in it, which was not even available as a preprint at that time.
I do not think I understood anything back then, as I was a high school student at that time.
But somehow, by a miraculous action at distance, I became a student of Eguchi-sensei several years later, and started specializing in the Seiberg-Witten theory.
It was Eguchi-sensei who advised me to read Nekrasov's influential instanton counting paper when it appeared in 2002, which became the basis of my career.

In the last year of the graduate school,  
recommendation letters by him and by Hirosi Ooguri allowed me
to stay in KITP from August to December 2005 for a long-term workshop.
Despite all his duties in the university, he also attended the workshop for one month in August.
While preparing for my first visit to the US, he suggested me to stay together with him in a large apartment already arranged for him by KITP, since he did not want to live alone.
It is a fond memory that I cooked barely edible breakfast for him every day, and often went to restaurants in Isla Vista together for supper.
He also taught me then what is according to him the most delicious and at the same time the easiest way to eat Avocado, 
which is to cut one in half, remove the big seed, and  then pour a bit of soy sauce to the resulting depression.
Then all that is left is to scoop the flesh by a spoon.
During my stay there, I worked with Suzuki-kun, who was another student of Eguchi-sensei, 
on 6d global anomaly cancellation across the Pacific Ocean using Skype.
As this paper can be considered as a direct sequel to that paper \cite{Suzuki:2005vu},
I feel it  quite appropriate that I dedicate this paper to his memory.

Due to the age difference, Eguchi-sensei was always a father-like figure to me,
including the following unfortunate events in his last few years.
In 2015, I was a faculty member in the department of physics of the University of Tokyo,
from which he had retired and moved to Rikkyo University.
Due to various reasons, I wanted to move to my current affiliation, 
and one day I went to Rikkyo University to visit him to ask if he could write a recommendation letter to me,
as he had done countless times before.
I was surprised then that he flatly refused to do so.
I think I now understand how he felt at that time; 
he had been in the same department of physics of the University of Tokyo for decades.
Surely he had felt strongly attached to the place, which I was suddenly trying to abandon.
Since then, we met once in a while in workshops and in other places,
but I never felt reconciled since,
just as a father and a son would feel when the partner brought back to the home by the son was not acceptable to the father.

In late 2018, I knew he was sick, but I was never able to make up my mind to visit him in the hospital.
Then came the sudden new of his passing away in January 30, 2019.
I deeply regret that I did not go to see him at least once, before he left this world.

On his passing, I recalled a parable of Zhuang-zi:\footnote{%
For the classical Chinese original and the modern English translation, see \url{https://ctext.org/zhuangzi/perfect-enjoyment\#n2831} .}\begin{quote}
莊子妻死，惠子弔之，莊子則方箕踞鼓盆而歌。惠子曰：「與人居長子，老身死，不哭亦足矣，又鼓盆而歌，不亦甚乎！」莊子曰：「不然。是其始死也，我獨何能無概然！察其始而本無生，非徒無生也，而本無形，非徒無形也，而本無氣。雜乎芒芴之間，變而有氣，氣變而有形，形變而有生，今又變而之死，是相與為春秋冬夏四時行也。人且偃然寢於巨室，而我噭噭然隨而哭之，自以為不通乎命，故止也。」
\end{quote}
Roughly: Zhuang-zi's wife died, and his friend Hui-zi came to mourn her. 
Hui-zi finds Zhuang-zi drumming and singing. 
Hui-zi says: ``Not mourning is one thing, but isn't drumming and singing a bit too much!''
Zhuang-zi replies: ``Not so. 
Of course I was sad when I first realized that my wife had died! 
But after reflecting how the life forms from nothingness and then goes back to nothingness,
it is no different from the change of four seasons.
She just started sleeping comfortably in a large room, and there was no use for me to cry loudly,
so I decided to stop.''

Eguchi-sensei is no more, and but in some sense his existence went back to become one with the world, or the universe, or whatever I should call this entirety, to which I will eventually return, too.

\def\arxivfont{\rm}
\bibliographystyle{ytamsalpha}
\baselineskip=.95\baselineskip
\bibliography{ref}

\end{document}